\documentclass[12pt]{iopart}

\usepackage{graphicx}
\usepackage{iopams}
\usepackage{wasysym}

\newcommand{\dif}{\mathrm{d}}
\newcommand{\expt}{\mathrm{e}}
\newcommand{\dfrac}{\frac}

\begin{document}

\title[Wave turbulence in preheating III]{Aspects of wave turbulence in preheating III: The case of the two-fields models}

\author{J. A. Crespo and H. P. de Oliveira}

\address{Universidade do Estado do Rio de Janeiro, Instituto de F{\'{\i}}sica - Departamento de F{\'{\i}}sica Te{\'o}rica, CEP 20550-013. Rio de Janeiro, RJ, Brazil.}

\ead{jaacrespo@gmail.com, hp.deoliveira@pq.cnpq.br}

\begin{abstract}
The present work is the continuation of the investigation of aspects of wave turbulence in preheating we have started in Refs. \cite{henrique13,crespo2019} but considering several classes of inflationary two-fields models. We exhibit the main elements of the wave turbulence phase imprinted in the power spectra of relevant quantities in the time and space domains. As a general feature, the power spectra resemble in their structure to those obtained using the single nonminimally coupled scalar field of the previous work. As a consequence, we obtained an estimate of the temperature corresponding to the thermalized phase from the power spectrum of the total energy density. Another feature allowed by the simulations was the determination of the effective equation of state after taking into account the backreaction of the produced particles.
\end{abstract}

%
%
\submitto{\JCAP}
%
\maketitle
%
%

\section{Introduction}

In the previous work \cite{crespo2019}, from now on called paper I, we studied the dynamics and the features of wave turbulence in the preheating model with a single scalar field nonminimally coupled to the spacetime curvature with the quartic self-interacting potential. We remark the motivation provided by the latest observational tests \cite{planck13,planck18} that showed the feasibility of this class of single field inflationary models. 

The wave turbulence is the primary mechanism for the energy transfer from the inflaton field to its inhomogeneous fluctuations \cite{henrique13,crespo2019,micha04,dufaux2006} during the nonlinear stages of preheating, and its main features are imprinted in the power spectra of relevant quantities in the time and the wavenumber. In particular, we have found that for the energy density, the corresponding power spectrum in wavenumber obtained in the turbulent regime reveals a component associated with the Planck distribution with a well-defined temperature identified as the reheating temperature. Moreover, in this phase, the resulting energy-momentum tensor considering the contribution of the inhomogeneous fluctuations (cf. Eq. 4 of the paper I) has a characteristic equation of state of a radiation fluid. 

An issue that deserves a throughout the investigation is whether or not the aspects of the wave turbulence would change considerably if distinct classes of models are taken into consideration. In this direction, we study here the dynamics of preheating of the inflationary two-field models already considered by Dufaux et al. \cite{dufaux2006}, one of them is the inflaton, $\phi$ coupled with another field $\chi$ as we are going to describe in the sequence. The structure of the present work does not differ from the paper I. Thus, in Section 2 we establish the basic equations for the nonlinear preheating in inflationary two-field models. We integrate the field equations taking into account the influence of the fluctuations into the expansion of the Universe with the code described in the previous paper. This is the subject of Section 3. In Section 4 we present the results after specifying three two-field models. Mainly, the results comprise the same topics of paper I, namely, the interaction between modes, the effect of backreaction of created particles to form an effective equation of state and power spectra in the time and the wavenumber. Finally, we conclude and trace out possible perspectives of both works.

\section{The model}%

Let us consider two interacting scalar fields, $\phi(\mathbf{x},t)$ and $\chi(\mathbf{x},t)$, represented by the Lagrangian  
\begin{equation}
\mathcal{L} = \sqrt{-g}\left[\dfrac{R}{2m_{pl}^2} - \dfrac{1}{2} \nabla_\mu \phi \nabla^\mu \phi - \dfrac{1}{2} \nabla_\mu \chi \nabla^\mu \chi - V\left(\phi,\chi\right)\right],\label{langrangeana_f}
\end{equation}

\noindent where $R$ is the scalar curvature, $\sqrt{-g}$ is the determinant of the spacetime metric, $m_{pl} = \left(8 \pi G \right)^{-1/2}$ is the reduced Planck mass and $V\left(\phi,\chi\right)$ is given by 
\begin{equation}
V\left(\phi,\chi\right) = \dfrac{m^2}{2}\phi^2 + \dfrac{\sigma_3}{2}\phi\chi^2+\dfrac{g_4^2}{2} \phi^2\chi^2+ \dfrac{\lambda_\chi}{4}\chi^4. \label{pot_34_legs}
\end{equation}

\noindent The potential (\ref{pot_34_legs}) was studied due to its appearance in many theoretical models and had been explored by Dufaux et al. \cite{dufaux2006}. 

The Einstein's field equations reads
\begin{equation}
\fl G_{\mu\nu} = \frac{1}{m_{pl}^2}\left\lbrace \partial_\mu \phi \partial_\nu \phi + \partial_\mu \chi \partial_\nu \chi - g_{\mu\nu} \left[\frac{1}{2}(\partial \phi)^2 + \frac{1}{2}(\partial \chi)^2 - V\left(\phi,\chi\right) \right]\right\rbrace,
\end{equation}

\noindent and the Klein-Gordon equations for the fields $\phi$ and $\chi$ are respectively

\begin{equation}
\square \phi - \frac{\partial V}{\partial \phi} = 0 \quad \mbox{and} \quad \square \chi - \frac{\partial V}{\partial \chi} = 0, \label{kg_eq0}
\end{equation}

\noindent where $\square \phi = g^{\mu\nu} \nabla_\mu(\partial_\nu \phi)$. 

The spacetime is described by the Friedmann-Lema{\^\i}tre-Robertson-Walker (FLRW) flat spacetime with the line element
\begin{equation}
\dif s^2 = - \dif t^2 + a^2\left(t\right) \dif \mathbf{x}. \dif \mathbf{x}, \label{metrica_flrw}
\end{equation}

\noindent where $a(t)$ is the scale factor. The inflaton scalar field $\phi(\mathbf{x},t)$ has a homogeneous component, $\phi_0(t)$ that plays a relevant role during the inflationary phase, and an inhomogeneous component, $\delta \phi(\mathbf{x},t)$ due the inflaton quantum fluctuations. The coupled scalar field $\chi(\mathbf{x},t)$ has a null homogeneous component. Thus,  only its fluctuations, $\delta \chi(\mathbf{x},t)$, enter in the dynamics and eventually are amplified during the initial stages of the preheating. Both scalar field fluctuations affect the dynamics of the spacetime according to the following form of the field equations %
\begin{equation}
G_{\mu\nu} = \frac{1}{m_{pl}^2} \left\langle T_{\mu\nu}\right\rangle, \label{einstein_eq}
\end{equation}

\noindent where $\left\langle ... \right\rangle$ denotes the spatial average in the physical domain and $T_{\mu\nu}$ is the energy-momentum of Eq. (3). We impose the condition $\left\langle \delta \phi(\mathbf{x},t) \right\rangle = 0$ and $\left\langle \delta \chi(\mathbf{x},t) \right\rangle = 0$ for the fluctuations.

We introduce dimensionless variables to write the field equations for the numerical integration. For the models under consideration, we have
\begin{equation}
\mathbf{x}_p = m\mathbf{x}, \quad t_p = m t, \quad \phi_p =  a^{3/2} \phi^{-1}_e \phi, \quad \chi_p =  a^{3/2} \phi^{-1}_e \chi. \label{adim_var_phi2}
\end{equation}

\noindent The subscript $p$ indicates the dimensionless version of the variable in the computational domain, $\phi_e$ is the amplitude of the homogeneous inflaton field at the end of inflation, i. e. $\phi_e=\phi_0(t_{\mathrm{end}})$, and $t_p \rightarrow t$ is the dimensionless time. With these new definitions, the Klein-Gordon equations (\ref{kg_eq0}) becomes 

\begin{eqnarray}
\phi''_p + \left[1 - \dfrac{3}{4}\left(\frac{a'}{a}\right)^2-\dfrac{3}{2}\dfrac{a''}{a}\right] \phi_p - \dfrac{\nabla^2_p \phi_p}{a^2} + \dfrac{q_3}{2}\dfrac{\chi^2_p}{a^{3/2}} + q_4 \dfrac{\chi_p^2 \phi_p}{a^3} & = & 0, \label{kg_phi_2}\\
\chi''_p - \left[\dfrac{3}{4}\left(\frac{a'}{a}\right)^2+\dfrac{3}{2}\dfrac{a''}{a}\right] \chi_p - \dfrac{\nabla^2_p \chi_p}{a^2} + q_3\dfrac{\chi_p \phi_p}{a^{3/2}} + q_4 \dfrac{\chi_p \phi_p^2}{a^3} + q_{\chi}\dfrac{\chi_p^3}{a^3} & = & 0, \label{kg_chi}
\end{eqnarray}

\noindent where
\begin{equation}
q_3 = \dfrac{\sigma_3\phi_e}{m^2}, \quad q_4 = \dfrac{g_4^2 \phi_e^2}{m^2}, \quad q_{\chi} = \dfrac{\lambda_\chi \phi_e^2}{m^2},
\end{equation}
\noindent $(...)' \equiv \partial(...)/ \partial t_p$ and $\nabla^2_p$ is the flat three-dimensional Laplacian operator. The remaining relevant field equations are

{\small
\begin{eqnarray}
\fl H^2_p &\fl \quad = &\fl \quad \quad \frac{m^2\phi^2_e}{3 m_{pl}^2a^3} \left\langle \frac{1}{2} \left[\left(\phi'_p - \dfrac{3}{2}H_p\phi_p\right)^2 + \left(\chi'_p - \dfrac{3}{2}H_p\chi_p\right)^2\right] +\frac{(\boldsymbol{\nabla}_p \phi_p)^2 + (\boldsymbol{\nabla}_p \chi_p)^2}{a^2} + V_p \right\rangle \label{eq_h_phi2} \\
\nonumber\\
\fl \dfrac{a''}{a} &\fl \quad = &\fl \quad \quad -\dfrac{\phi_e^2}{3m_{pl}^2 a^3}\left\langle - \left(\phi'_p - \dfrac{3}{2}H_p\phi_p\right)^2 - \left(\chi'_p - \dfrac{3}{2}H_p \chi_p\right)^2 + \frac{(\boldsymbol{\nabla}_p \phi_p)^2 + (\boldsymbol{\nabla}_p \chi_p)^2}{a^2} + V_p \right\rangle \label{eq_a2_phi2}
\end{eqnarray}
}
\noindent where $H_p = a'/a$ and

\begin{equation}
V_p = \frac{1}{2}\phi_p^2 + \frac{q_3}{2a^{3/2}}\phi_p \chi_p^2 + \frac{q_4}{2a^3}\phi_p^2 \chi_p^2 + \frac{q_\chi}{4a^3}\chi_p^4.
\end{equation}

\noindent We integrate the system of field equations (\ref{kg_phi_2}), (\ref{kg_chi}) and (\ref{eq_a2_phi2}) numerically  starting at the end of inflation and using the same numerical scheme of paper I and briefly described in the next Section.


\section{Numerical approach}

As in the paper I, we integrate the field equations (\ref{kg_phi_2}), (\ref{kg_chi}) and (\ref{eq_a2_phi2}) in a three-dimensional square box of comoving size $L$ with periodic boundary conditions as a suitable model of the Universe. We described the algorithm with details in the precedent paper, and here we present the spectral approximations for both fields and the final form of the field equations for the sake of completeness. 

We establish the following spectral approximations for the scalar fields $\phi$ and $\chi$: 

\begin{eqnarray}
\phi_N(\mathbf{x}_p,t) &&= \alpha_\mathbf{0}(t) +  \sum_{\mathbf{k} \neq \mathbf{0}}\widehat{\phi}_\mathbf{k}(t) \psi_\mathbf{k}(\mathbf{x}_p) \\
\nonumber \\
\chi_N(\mathbf{x}_p,t) &&= \sum_{\mathbf{k} \neq \mathbf{0}}\widehat{\chi}_\mathbf{k}(t) \psi_\mathbf{k}(\mathbf{x}_p).
\end{eqnarray} 

\noindent Here $\mathbf{k}=(k_x,k_y,k_z)$ is the comoving momentum where each component varies from $-N/2$ to $N/2-1$ and $N$ is the truncation order. The Fourier basis functions are $\psi_\mathbf{k}(\mathbf{x}_p)=\exp\left(\frac{2\pi i}{L_p}\mathbf{k}\cdot\mathbf{x}_p\right)$ with $L_p$ being the dimensionless box size. The unknown modes $\widehat{\phi}_\mathbf{k}(t)$ and $\widehat{\chi}_\mathbf{k}(t)$ are complex functions that satisfy the relations $\widehat{\phi}_\mathbf{k}(t)=\widehat{\phi}^*_\mathbf{-k}(t)$ and $\widehat{\chi}_\mathbf{k}(t)=\widehat{\chi}^*_\mathbf{-k}(t)$ to produce real scalar fields.  In this context, $\alpha_\mathbf{0}(t)$ is the homogeneous component of the inflaton field and the corresponding homogeneous component of $\chi$ is absent, or $\widehat{\chi}_\mathbf{0}(t)=0$.
	
Following the procedure delineated in the previous paper, we substitute the approximations (14) and (15) into the Klein-Gordon equations (\ref{kg_phi_2}) and (\ref{kg_chi}) to form the corresponding residual equations. According to the pseudospectral method, the unknown modes are such that the residuals vanish at the collocation or grid points $\mathbf{x}_{p\mathbf{n}}=L_p\mathbf{n}/N$. Therefore, we arrive at
\begin{eqnarray}
\fl \mathrm{Res}_\phi(\mathbf{x}_{p\mathbf{n}},t) &\fl \quad \qquad \qquad \ = &\fl \qquad \qquad \qquad \phi^{\prime\prime}_{p\mathbf{n}} + \left[1 - \dfrac{3}{4}\left(\frac{a'}{a}\right)^2-\dfrac{3}{2}\dfrac{a''}{a}\right] \phi_{p\mathbf{n}} + \frac{4\pi}{L_p^2} \sum \vert\mathbf{k}\vert^2 \widehat{\phi}_{\mathbf{k}}(t) \psi_{\mathbf{k}}(\mathbf{x}_{p\mathbf{n}}) + \nonumber \\
 & & + \dfrac{q_3}{2}\dfrac{\chi^2_{p\mathbf{n}}}{a^{3/2}} + q_4 \dfrac{\chi_{p\mathbf{n}}^2 \phi_{p\mathbf{n}}}{a^3}=0, \label{sistema_edo1} \\
  & & \nonumber \\
\fl \mathrm{Res}_\chi(\mathbf{x}_{p\mathbf{n}},t) &\fl \quad \qquad \qquad \  = &\fl \qquad \qquad \qquad  \chi^{\prime\prime}_{p\mathbf{n}} - \left[\dfrac{3}{4}\left(\frac{a'}{a}\right)^2+\dfrac{3}{2}\dfrac{a''}{a}\right] \chi_{p\mathbf{n}}  + \frac{4\pi}{L_p^2} \sum \vert\mathbf{k}\vert^2 \widehat{\chi}_{\mathbf{k}}(t) \psi_{\mathbf{k}}(\mathbf{x}_{p\mathbf{n}}) + \nonumber \\
 & & + q_3\dfrac{\chi_{p\mathbf{n}} \phi_{p\mathbf{n}}}{a^{3/2}} + q_4 \dfrac{\chi_{p\mathbf{n}} \phi_{p\mathbf{n}}^2}{a^3} + q_{\chi}\dfrac{\chi_{p\mathbf{n}}^3}{a^3}=0. \label{sistema_edo2}
\end{eqnarray}
\noindent In the above expressions, we express the Laplacian term in function of the modes; $\phi_{p\mathbf{n}}$ and $\chi_{p\mathbf{n}}$ are the values of the scalar fields at the collocation points connected with the modes through 
\begin{eqnarray}
\phi_{p\mathbf{n}} & = & \phi_N(\mathbf{x}_{p\mathbf{n}},t)=\sum \widehat{\phi}_{\mathbf{k}}(t) \psi_{\mathbf{k}}(\mathbf{x}_{p\mathbf{n}}). \label{phi_expr} \\
\chi_{p\mathbf{n}} & = & \chi_N(\mathbf{x}_{p\mathbf{n}},t)=\sum \widehat{\chi}_{\mathbf{k}}(t) \psi_{\mathbf{k}}(\mathbf{x}_{p\mathbf{n}}), \label{chi_expr}
\end{eqnarray}
\noindent where we use the fast Fourier transform (FFT) to invert this relation whenever necessary.

The structure of the Eqs. (16) and (17) shows that some quantities are expressed in terms of the modes $\widehat{\phi}_{\mathbf{k}}(t),\widehat{\chi}_{\mathbf{k}}(t)$ while other in terms of the values $\phi_{p\mathbf{n}},\chi_{p\mathbf{n}}$ wherever is more convenient \cite{orszag}. We take advantage of this strategy to calculate the spatial average (cf. Eq. (19) of \cite{crespo2019}) of some quantities appearing in Eqs. (11) and (12) without any approximation:
\begin{eqnarray}
\left\langle \phi_p^2 \right\rangle & = & \sum\limits_\mathbf{k}\left\vert \widehat{\phi}_{\mathbf{k}}\right\vert^2, \\
\left\langle \phi_p^{\prime 2} \right\rangle & = & \sum\limits_\mathbf{k}\left\vert \widehat{\phi}'_{\mathbf{k}}\right\vert^2, \\
\left\langle \phi'_p \phi_p \right\rangle & = & \sum\limits_\mathbf{k} \widehat{\phi}'_{\mathbf{k}}\widehat{\phi}_{\mathbf{k}}^{*}, \\
\left\langle \left(\boldsymbol{\nabla}_p\phi_p\right)^2 \right\rangle & = & \dfrac{4\pi^2}{L_p^2}\sum\limits_\mathbf{k} k^2 \left\vert \widehat{\phi}_{\mathbf{k}}\right\vert^2,
\end{eqnarray}

\noindent where $\widehat{\phi}_{p\mathbf{k}}^{*}$ is the complex conjugate of $\widehat{\phi}_{p\mathbf{k}}$. The same approach can be used for $\chi_p$ terms. The spatial average of $\chi_p^4$ term was the only one numerically evaluated using quadrature formula.
 
The integration starts at the end of inflation with the following initial conditions $\alpha_{\mathbf{0}}(0) = 0$, $\alpha^\prime_{\mathbf{0}}(0) = 1$, $a(0)=1$ and the initial modes and their first derivatives, $\widehat{\phi}_{\mathbf{k}}(0)$,$\widehat{\chi}_{\mathbf{k}}(0)$, $\widehat{\phi}^\prime_{\mathbf{k}}(0)$, $\widehat{\chi}^\prime_{\mathbf{k}}(0)$, respectively, have typical amplitudes of the order of $10^{-4}$. We employ the FFT to obtain the corresponding values $\phi_{p\mathbf{n}}(0)$, $\chi_{p\mathbf{n}}(0)$ as well their first derivaties which are nedeed for the equations (11), (12), (16) and (17). Finally, the initial $a^\prime(0)$ can be calculated from the Friedmann equation (11). Then, with these initial values we can determine $a, a^\prime, \widehat{\phi}_{\mathbf{k}},\widehat{\chi}_{\mathbf{k}},\widehat{\phi}^\prime_{\mathbf{k}},\widehat{\phi}^\prime_{\mathbf{k}}$ at the next time level (here the FFT is used to update the modes associated to both scalar fieids) and the whole process repeats resulting in the time march. We adopted the fourth order Runge-Kutta integrator using a stepsize h = 0.001.

\section{Wave turbulence towards the thermalization}

We investigate the influence of different interactions on the aspects of wave turbulence developed in the nonlinear stage of preheating. These of models were well explored in the preheating context \cite{micha04,kofman94,prokopec97,greene97,kofman97,kofman98,podolski06}, mainly with the 4-legs potential 
\begin{equation}
V_4\left(\phi,\chi\right) = \dfrac{1}{2}g_4^2\phi^2\chi^2, \label{pot_4-legs}
\end{equation} 

\noindent where $g_4$ is the coupling constant, $\phi$ is the inflaton field, and $\chi$ represents another material scalar field excited by the decay of $\phi$. However, this decay is not complete when only this kind of interaction is present, behavior necessary for the radiation era of the Universe \cite{podolski06}. To circumvent this problem the following 3-legs interaction is introduced 
\begin{equation}
V_3\left(\phi,\chi\right) = \dfrac{1}{2}\sigma_3\phi\chi^2. \label{pot_3-legs}
\end{equation} 

\noindent This kind of interaction naturally emerges in some theoretical models, as supersymmetric theories. These models have features similar to the tachionic preheating \cite{kofman2001}, leading the system to a ``tachionic resonance'' phase \cite{dufaux2006,podolski06}. Although the recent observational data \cite{planck13,planck2016} ruled out single field inflationary models based in potentials like $m^2\phi^2$, Harigaya et al. \cite{harigaya2016} have shown that including other scalar fields interacting with the inflaton results in models in agreement with the observational data.

We extend the analysis of \cite{henrique13} to include the effect of backreaction of the scalar field fluctuations in three distinct configurations \cite{dufaux2006}. The model A is the well-known two fields model where the resonance is driven only by the 4-legs interaction; the model B comes from supersymmetric theories with the presence of 3-legs and 4-legs interactions, and the model C has just the 3-legs interaction so that we can analyze the tachionic resonance effects. The self-interaction $\chi^4$ term is present in all models to guarantee the stability of the models with the presence of 3-legs potential \cite{dufaux2006}.

In all numerical experiments, we set $N=64$ that corresponds to a three-dimensional grid with $64^3$ points. For the sake of numerical resolution, we evolved the system in a total of $2^{20}$ time steps or a time interval of $\Delta t = 1048.576$. Also, we set the coupling constants as showed in Table \ref{q_values}  (cf. ref. \cite{dufaux2006}).
\begin{table}[ht]
  \centering
  \begin{tabular}{c c c c}
    \hline
      \textbf{Model} & $q_3$  & $q_4$  & $q_\chi$ \\
    \hline
      A      & $0$    & $10^4$ & $5 \times 10^3$\\
      B      & $10^2$ & $10^4$ & $5 \times 10^3$\\
      C      & $10^2$ & $0$    & $10^4$ \\
    \hline
  \end{tabular}
  \caption{Values of $q_3$, $q_4$ e $q_\chi$ for each model.}\label{q_values}
\end{table}

In two-fields models, the linear phase of preheating is governed by the Mathieu equation when only one interaction potential is present and governed by the Hill equation when 3-legs and 4-legs interactions are present \cite{dufaux2006}. In all cases we have a broad parametric resonant window and energy transfer occurs in a fast and efficient way in  the initial stage of preheating. There is no a specific value for $L_p$ and we choose $L_p = 10$. 

\subsection{Interaction between the modes}

Following a similar approach that we presented in the previous paper \cite{crespo2019} to study the energy transfer through different scales and the establishment of turbulent stages of the system, we analyze the decay of the homogeneous mode  $\alpha_\mathbf{0}(t)$ of the inflaton field and the variance of this scalar field $\sigma^2_\phi(\tau)$ given by

\begin{equation}
\sigma^2_\phi = \left\langle\left(\phi_p - \left\langle\phi_p\right\rangle\right)^2\right\rangle = \sum\limits_\mathbf{k}\left\vert \widehat{\phi}_{\mathbf{k}}\right\vert^2 - \alpha_\mathbf{0}^2. \label{var_phi}
\end{equation}

\noindent This quantity gives us a piece of relevant information about the contribution of the inhomogeneous modes of the inflaton field throughout the time evolution.
\begin{figure}[h!]
\includegraphics[scale=.58]{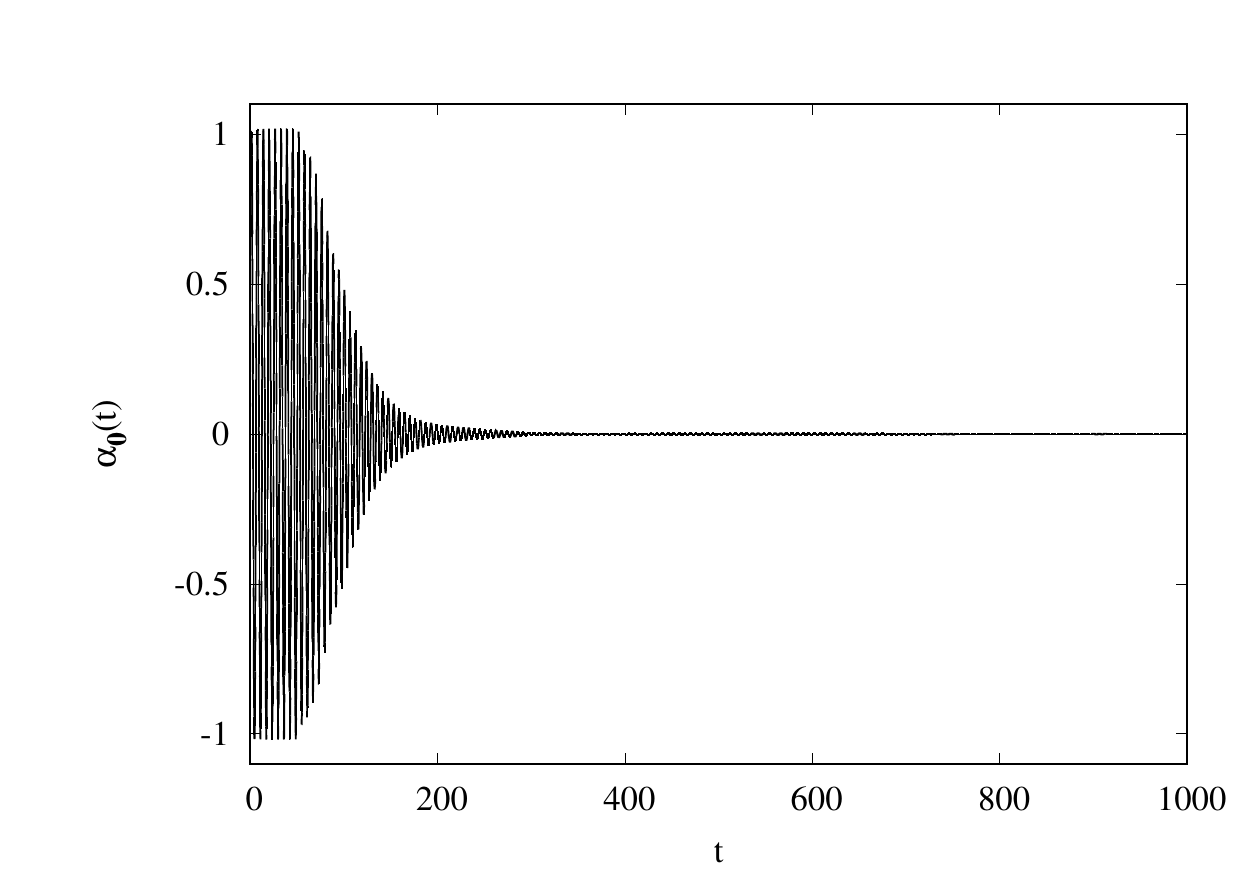}
\includegraphics[scale=.58]{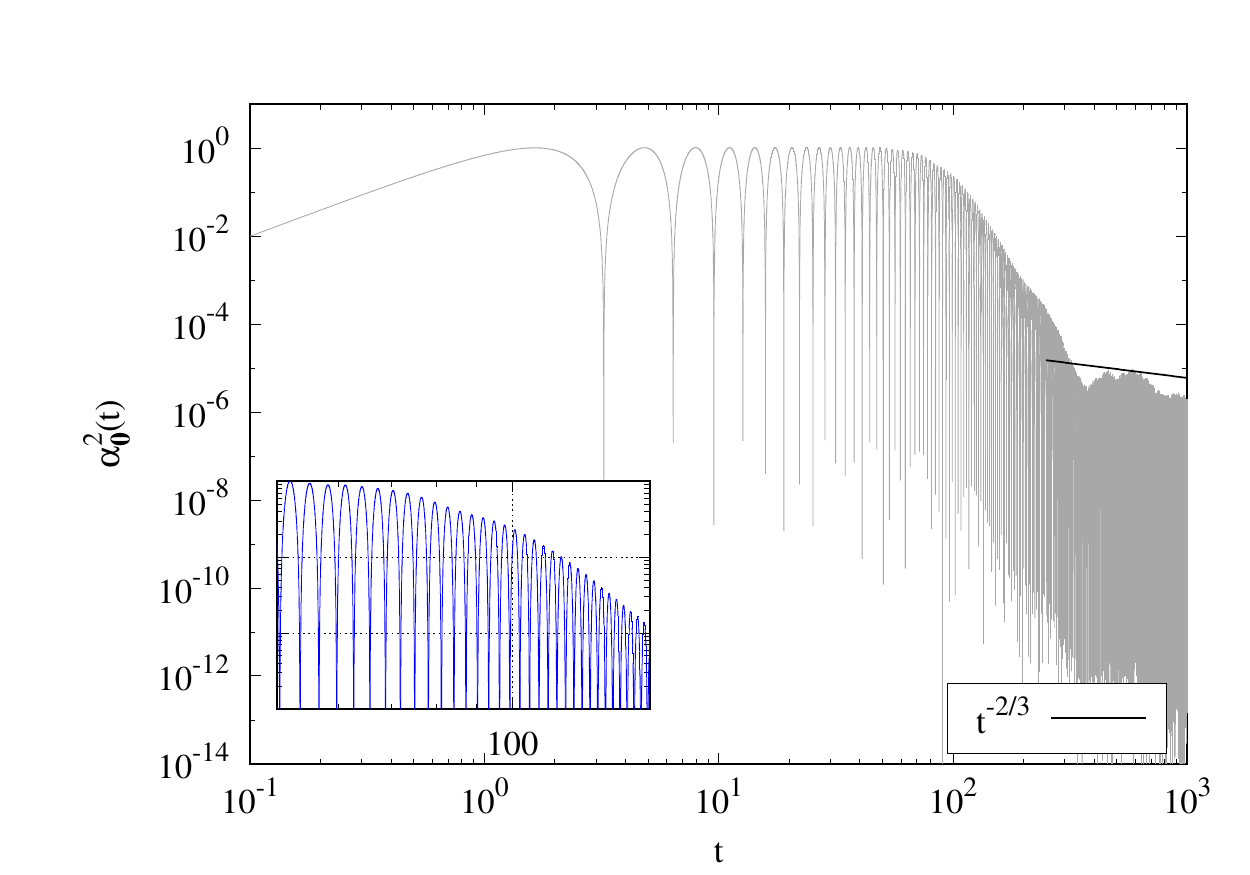}
\caption{\label{hom_A}Time evolution of the homogeneous mode, model A.}
\end{figure}
\begin{figure}[h!]
\includegraphics[scale=.58]{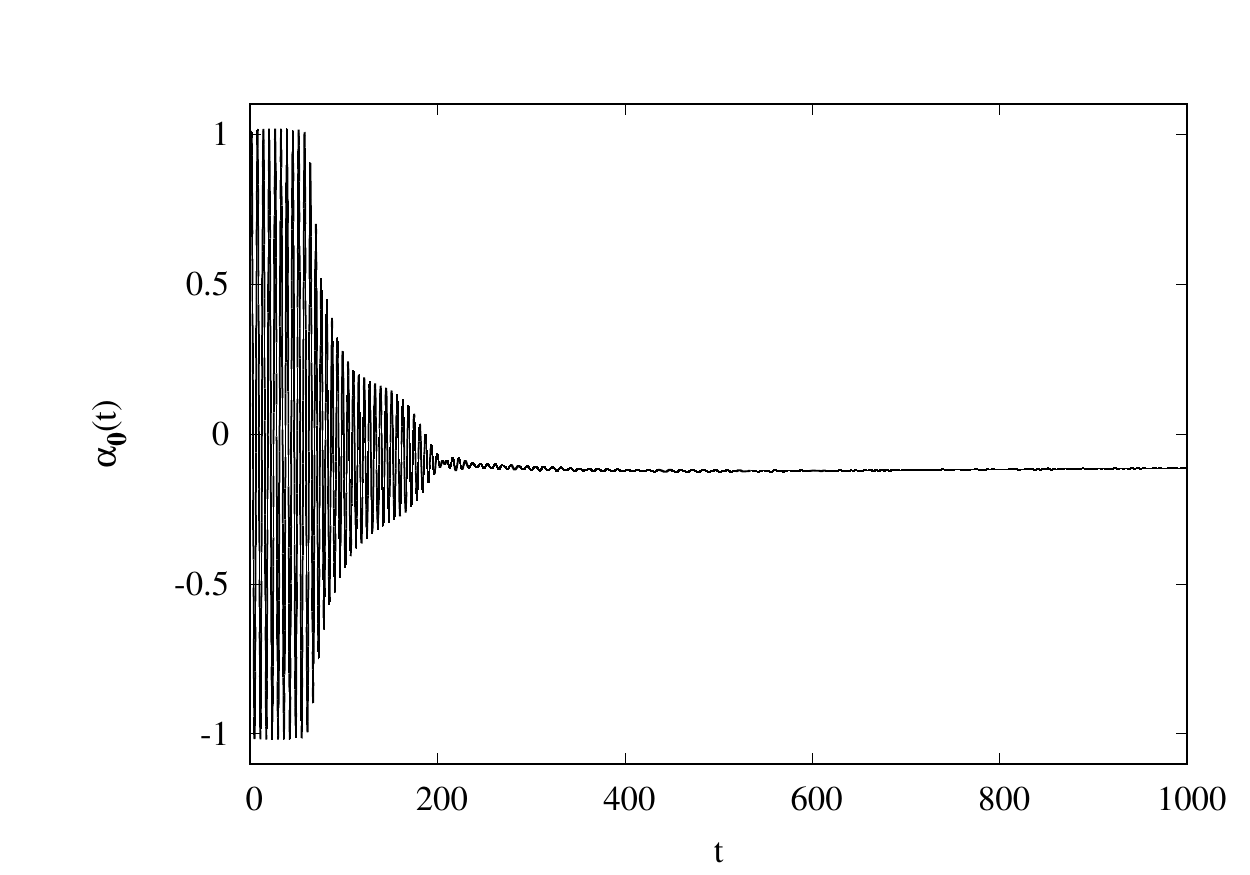}
\includegraphics[scale=.58]{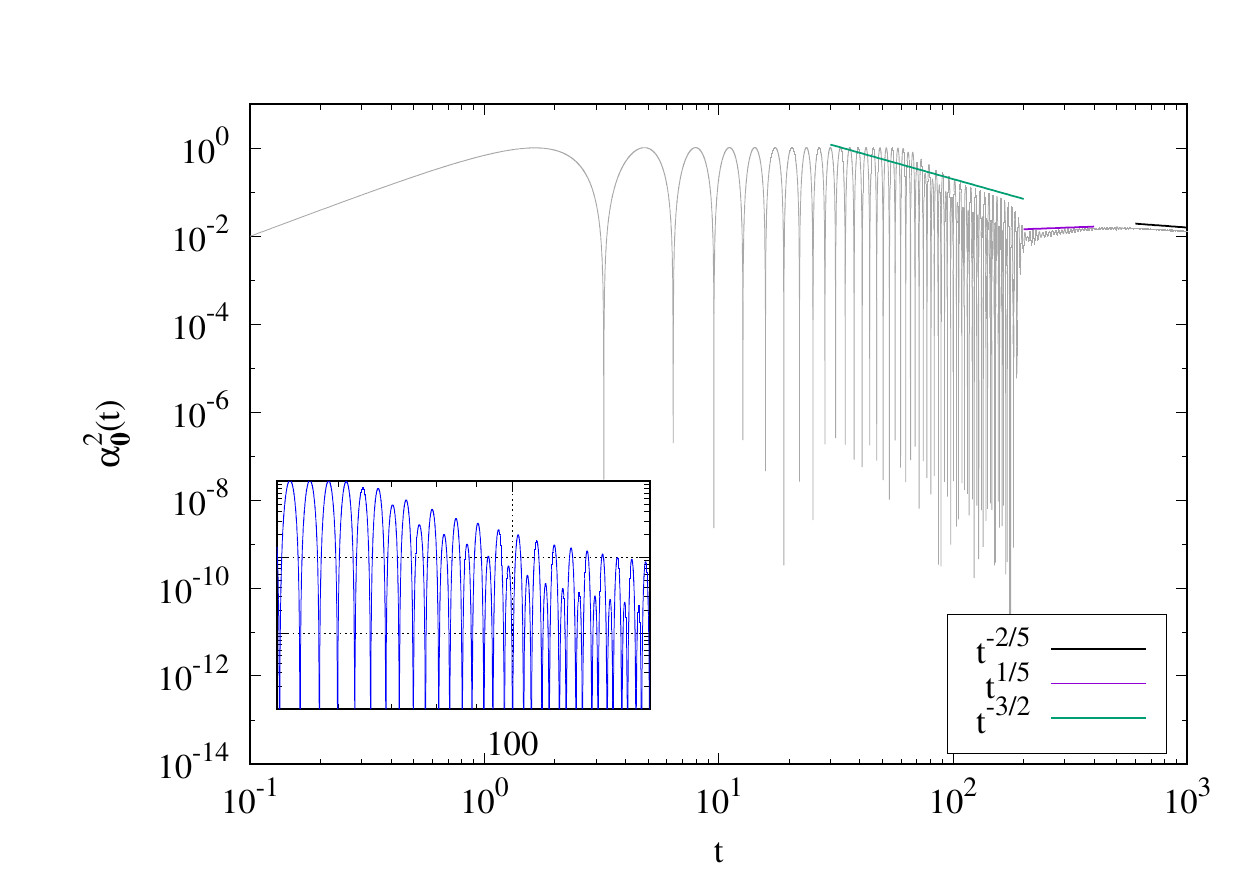}
\caption{\label{hom_B}Time evolution of the homogeneous mode, model B.}
\end{figure}
\begin{figure}[h!]
\includegraphics[scale=.58]{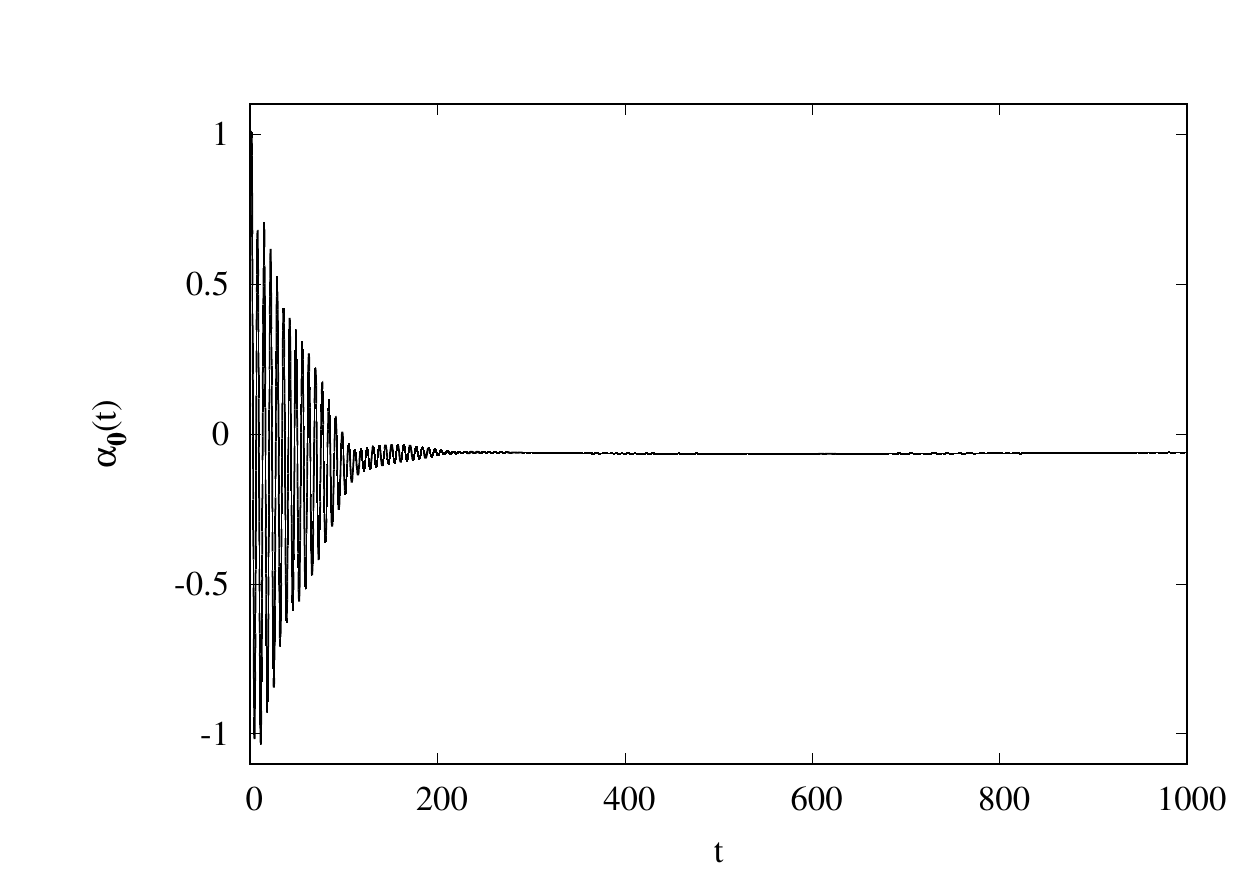}
\includegraphics[scale=.58]{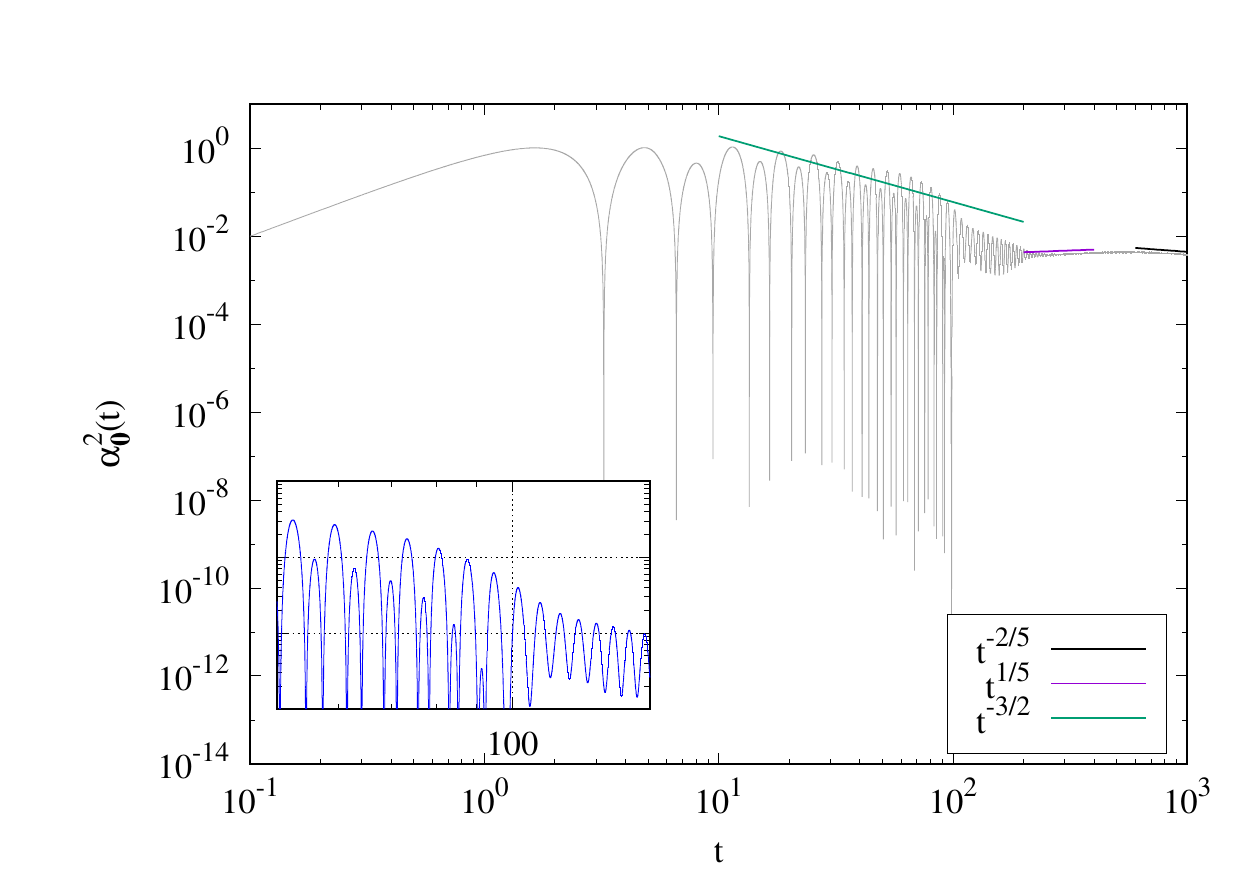}
\caption{\label{hom_C}Time evolution of the homogeneous mode, model C.}
\end{figure}

We notice the distinct influences of the interactions on the dynamics of the fields and the consequent energy transfer. Models A and B have the presence of 4-leg interaction, and until $t = 100$ (Figs. 1 and 2), they present a constant amplitude oscillation for $\alpha_\mathbf{0}$, which characterizes the linear phase of preheating. The model C has only the 3-leg interaction, and the duration of the linear stage of preheating is very short, ending after few oscillations of the homogeneous inflaton field, namely, it lasts until $t \approx 10$ (Fig. 3). After the decay of the amplitude of the inflaton oscillations, we notice that $\alpha_\mathbf{0}$ oscillates around a negative plateau in the models with the 3-legs interaction \cite{dufaux2006}. The homogeneous inflaton mode has similar behavior in the models A and B until $t \approx 200$. After this instant, its behavior changes, and the models B and C become very similar. We can infer that the 4-legs interaction dominates the model B in the linear phase, and 3-legs interaction becomes predominant in the subsequent nonlinear phase.

The inflaton variance has well-defined behavior in each model. Model A presents a modulated growth typical of broad parametric resonance followed by a decreasing and growth scaled by $t^{0.1}$. The model C has a resonance peak at right before $t \approx 10$, followed by an increase in $10\leqslant t \leqslant 100$ and decay scaled by $t^{-1/5}$.  Both behaviors are very similar those found in \cite{micha04} but we have to emphasize that we considered a massive inflaton with potential presented in \cite{dufaux2006} while \cite{micha04} adopted a self interacting massless inflaton with 4-legs interaction. The model B has an intermediary behavior between models A and C, and exhibits the same scale law of C.
\begin{figure}[h!]
\includegraphics[scale=.67]{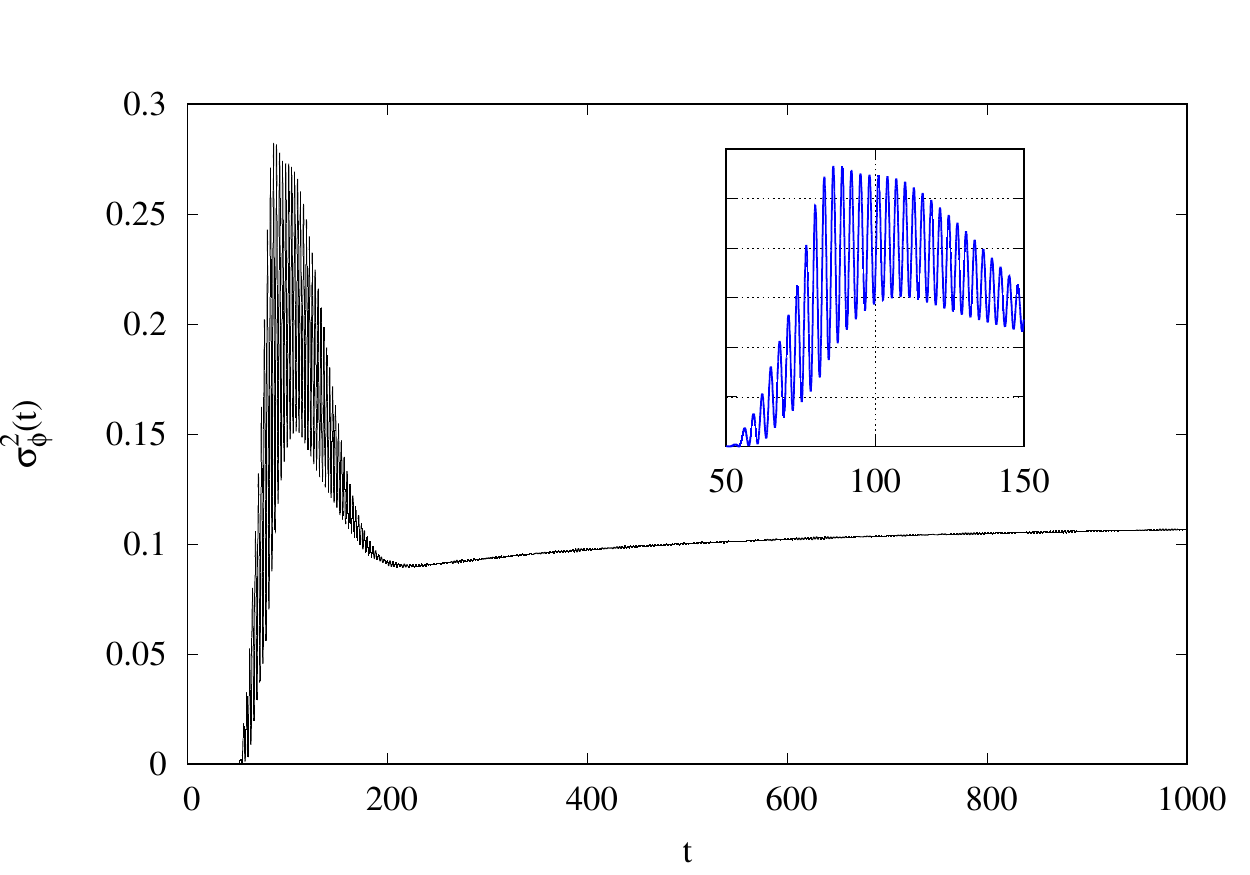}
\includegraphics[scale=.67]{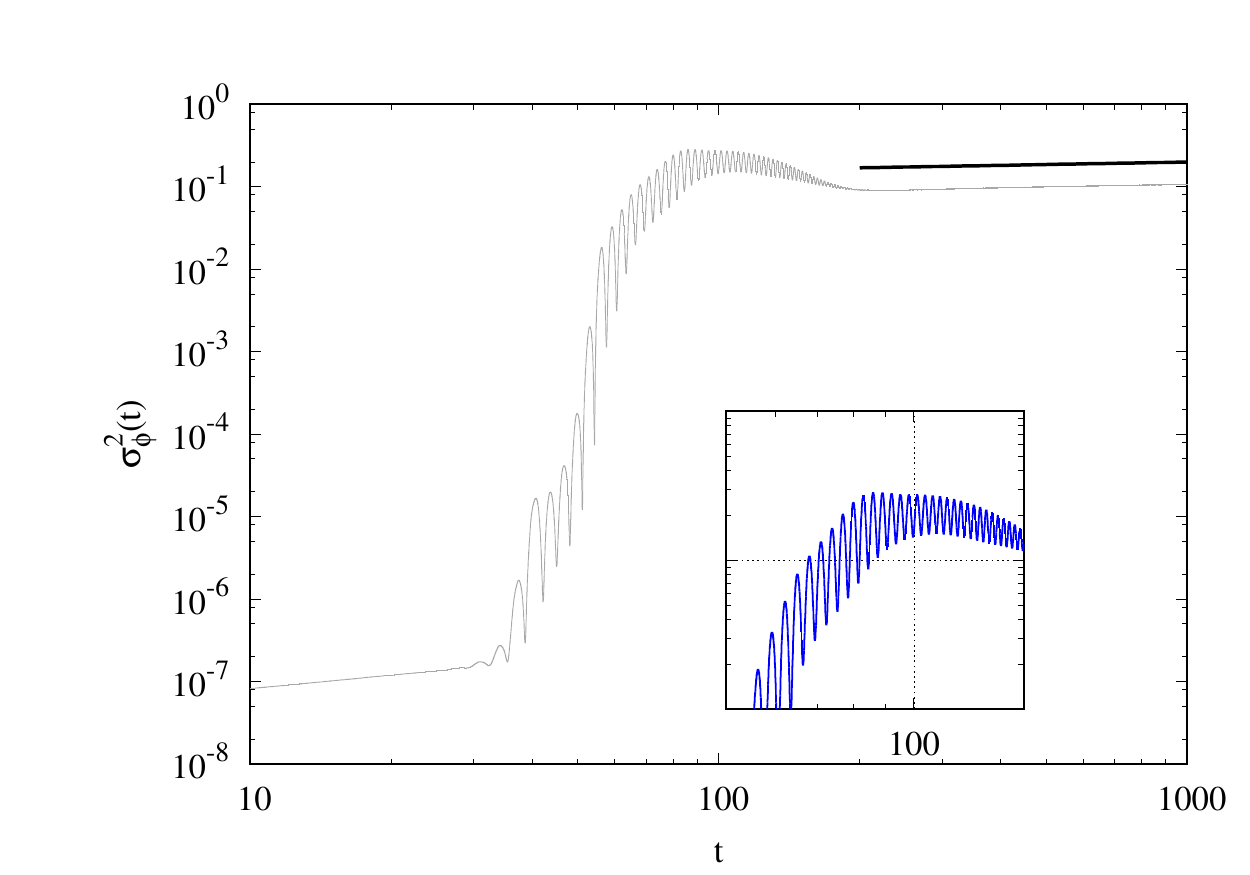}
\caption{\label{var_0} Time evolution of the inflaton variance, model A.}
\end{figure}
\begin{figure}[h!]
\includegraphics[scale=.67]{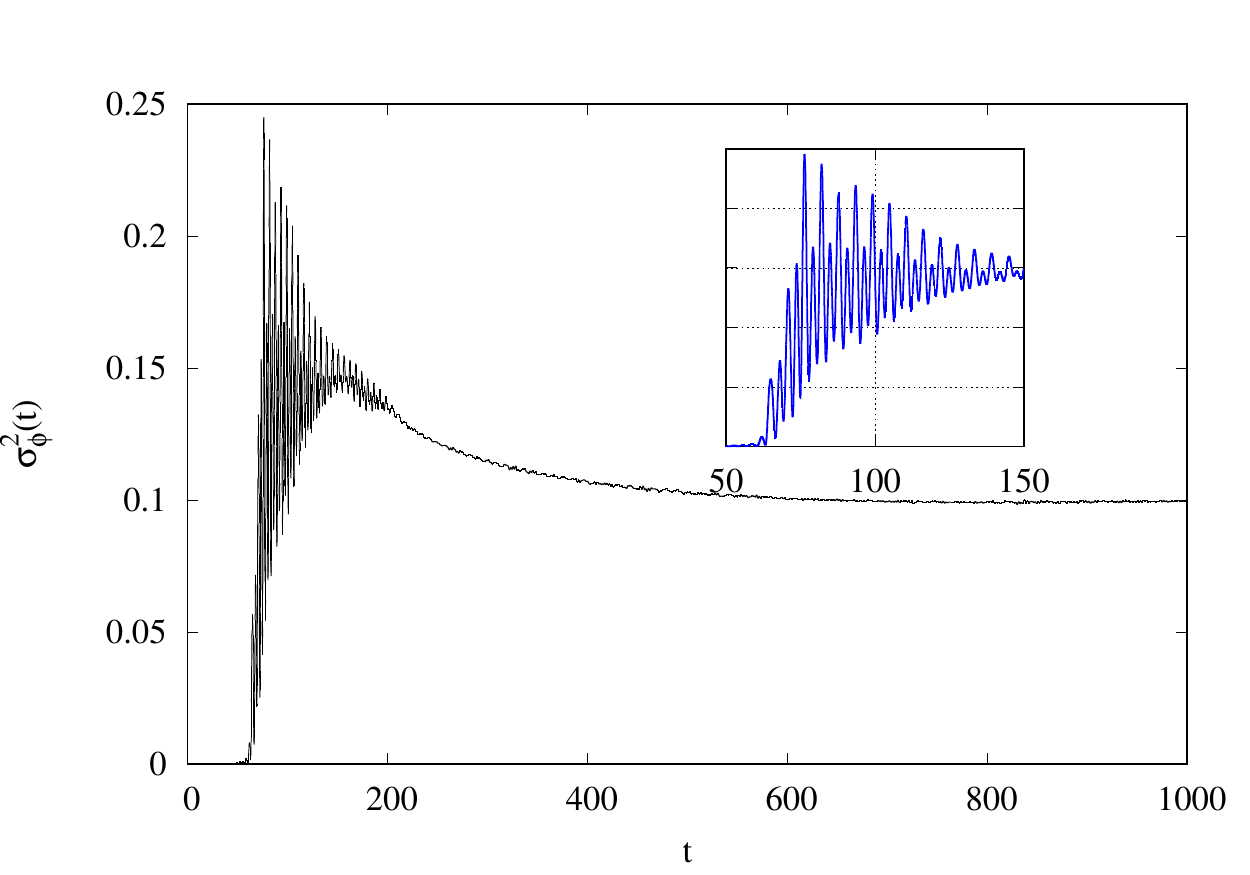}
\includegraphics[scale=.67]{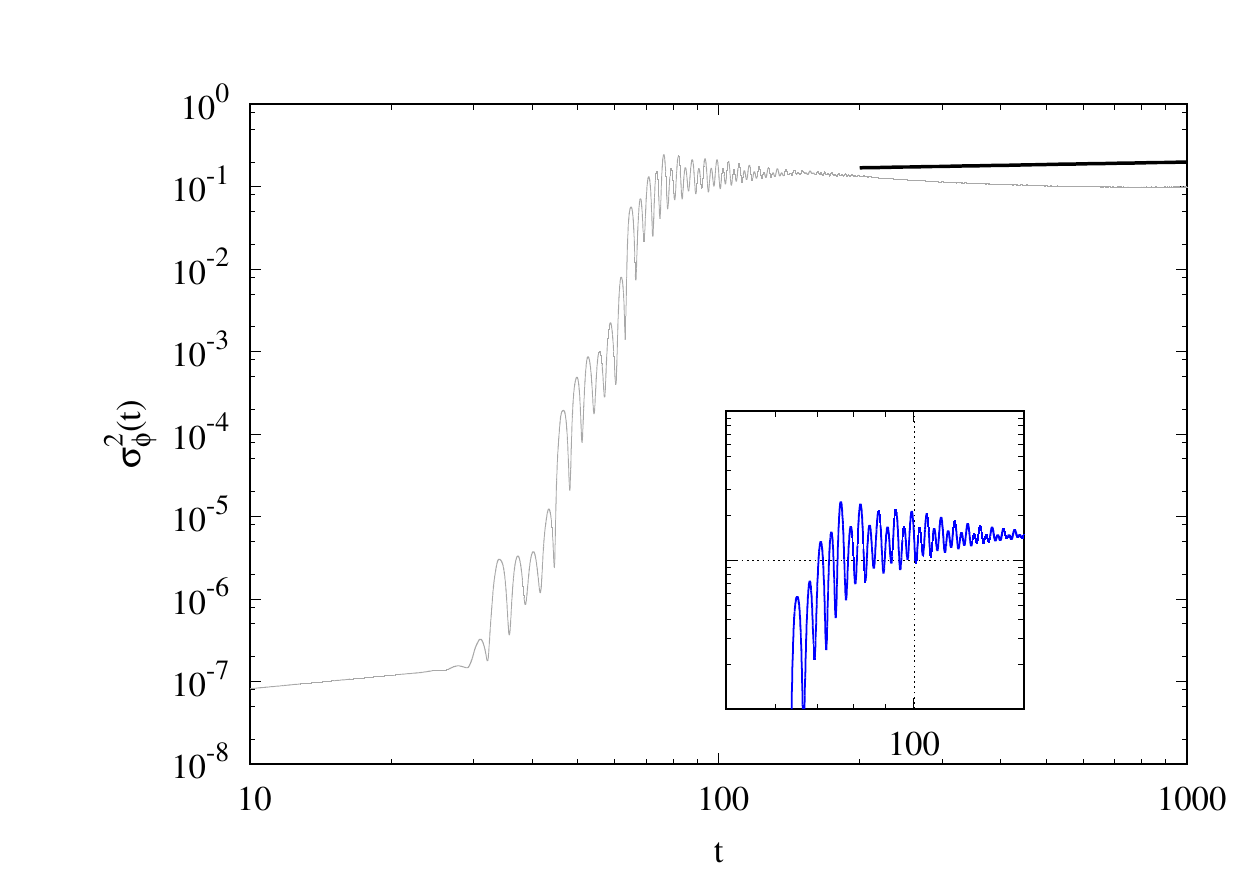}
\caption{\label{var_01} Time evolution of the inflaton variance, model B.}
\end{figure}
\begin{figure}[h!]
\includegraphics[scale=.67]{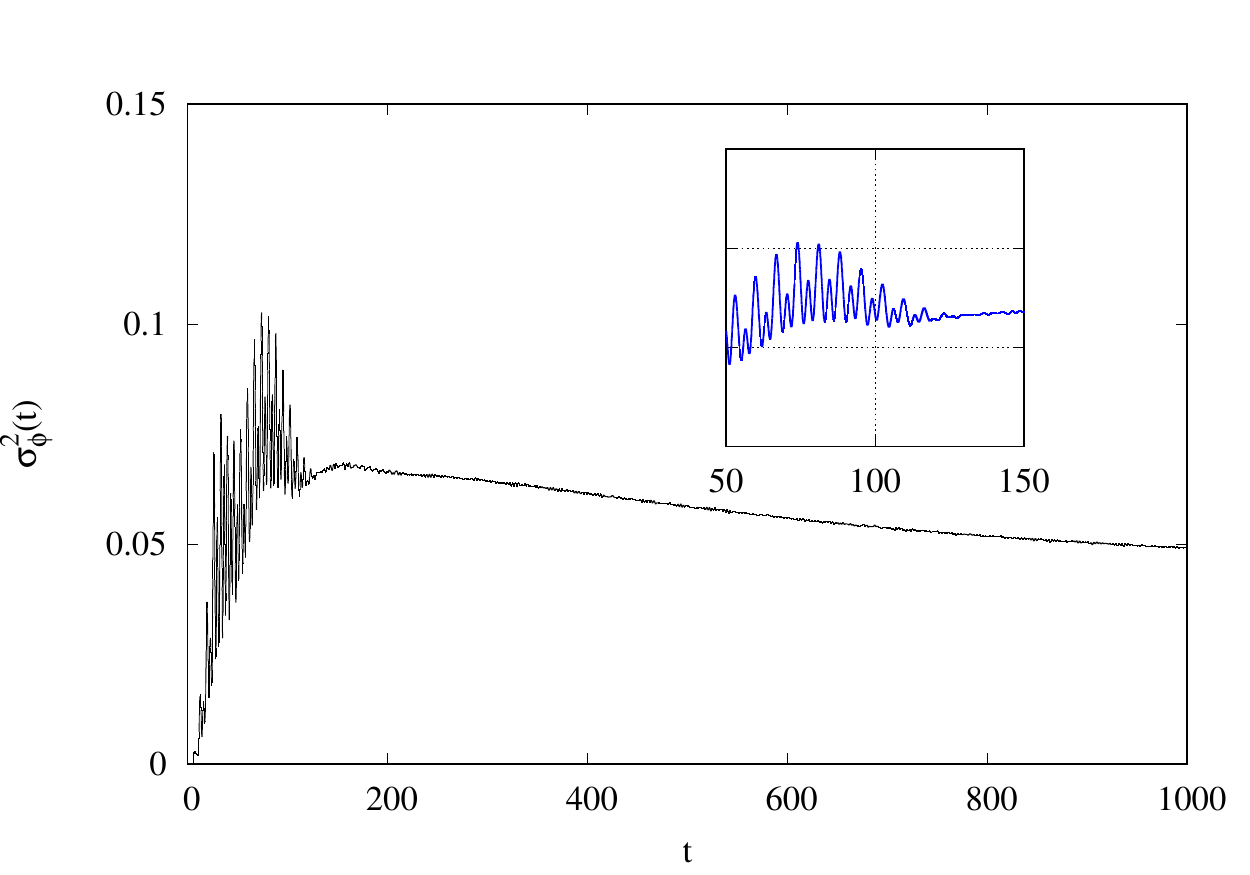}
\includegraphics[scale=.67]{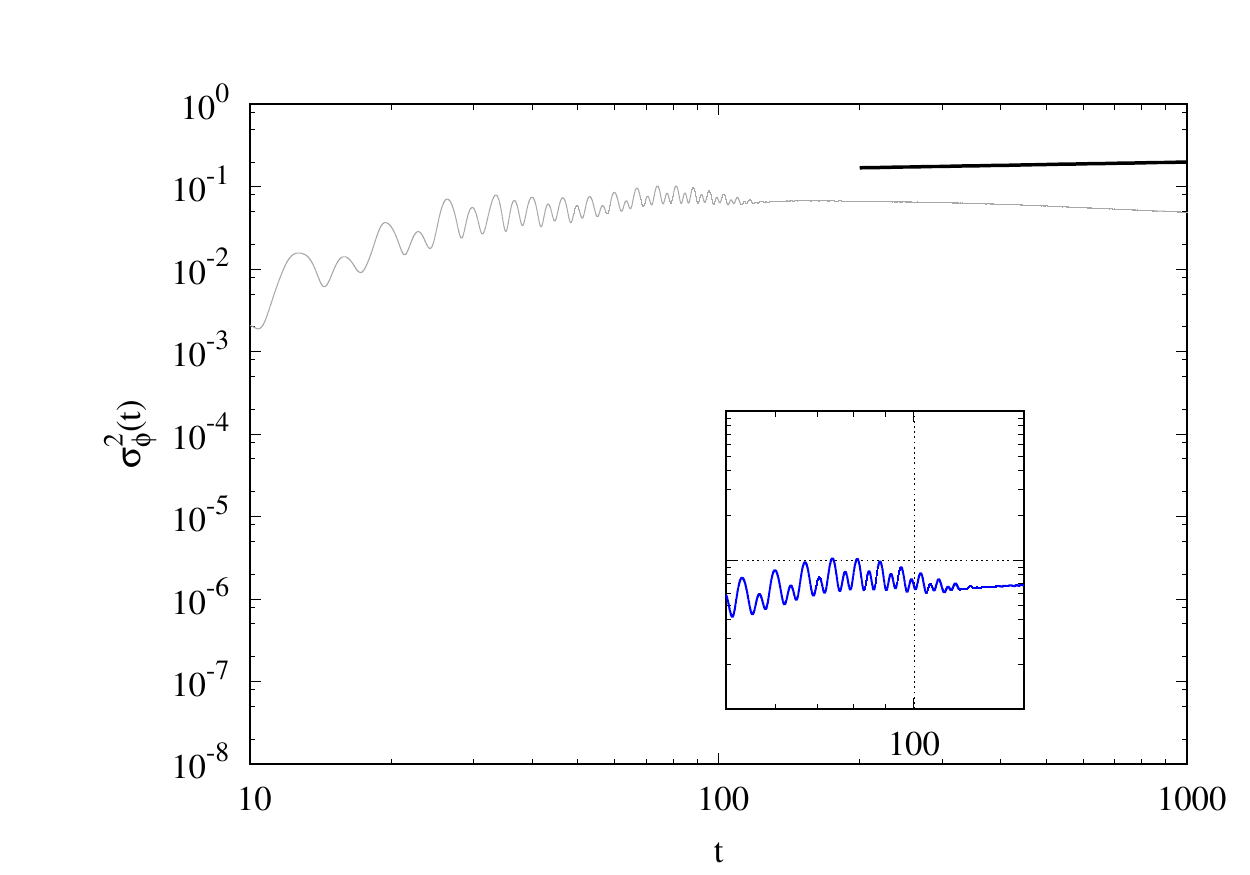}
\caption{\label{var_1} Time evolution of the inflaton variance, model C.}
\end{figure}

\subsection{Backreaction effects}

We proceed by showing the influence of different interactions on the backreaction of the created particles in the effective equation of state and the expansion of the universe described by the scale factor $a(t)$. The parameter $w$ that defines the equation of state is given by 

\begin{equation}
w\left(t\right) = \dfrac{\left\langle p \right\rangle}{\left\langle \rho \right\rangle}, \label{eos}
\end{equation}
\noindent where $\left< p \right>$ and $\left< \rho \right>$ are the effective  pressure and energy density, respectively with the corresponding expressions derived from the field equations (\ref{einstein_eq}):
\begin{eqnarray}
\rho & = & \dfrac{m^2\phi_0^2}{a^3}\left[ \dfrac{1}{2}\left(\phi'_p - \dfrac{3}{2}H_p \phi_p\right)^2 + \dfrac{1}{2}\left(\chi'_p - \dfrac{3}{2}H_p \chi_p\right)^2 + \dfrac{1}{2a^2}\left(\boldsymbol{\nabla}_p\phi_p\right)^2 +  \right. \nonumber\\
     &   & \left. + \dfrac{1}{2a^2}\left(\boldsymbol{\nabla}_p\chi_p\right)^2 + \dfrac{1}{2}\phi_p^2 + \dfrac{q_3}{2}\dfrac{\phi_p \chi_p^2}{a^{3/2}} + \dfrac{q_4}{2}\dfrac{\phi_p^2 \chi_p^2}{a^3} + \dfrac{q_\chi}{4} \dfrac{\chi_p^4}{a^3} \right], \\
     \nonumber \\
p & = & \dfrac{m^2\phi_0^2}{a^3}\left[ \dfrac{1}{2}\left(\phi'_p - \dfrac{3}{2}H_p \phi_p\right)^2 + \dfrac{1}{2}\left(\chi'_p - \dfrac{3}{2}H_p \chi_p\right)^2 - \dfrac{1}{6a^2}\left(\boldsymbol{\nabla}_p\phi_p\right)^2 - \right. \nonumber \\
  &   & \left. - \dfrac{1}{6a^2}\left(\boldsymbol{\nabla}_p\chi_p\right)^2 - \dfrac{1}{2}\phi_p^2 - \dfrac{q_3}{2}\dfrac{\phi_p \chi_p^2}{a^{3/2}} - \dfrac{q_4}{2}\dfrac{\phi_p^2 \chi_p^2}{a^3} - \dfrac{q_\chi}{4} \dfrac{\chi_p^4}{a^3} \right].
\end{eqnarray}

\begin{figure}[h]
\begin{center}
\includegraphics[scale=.61]{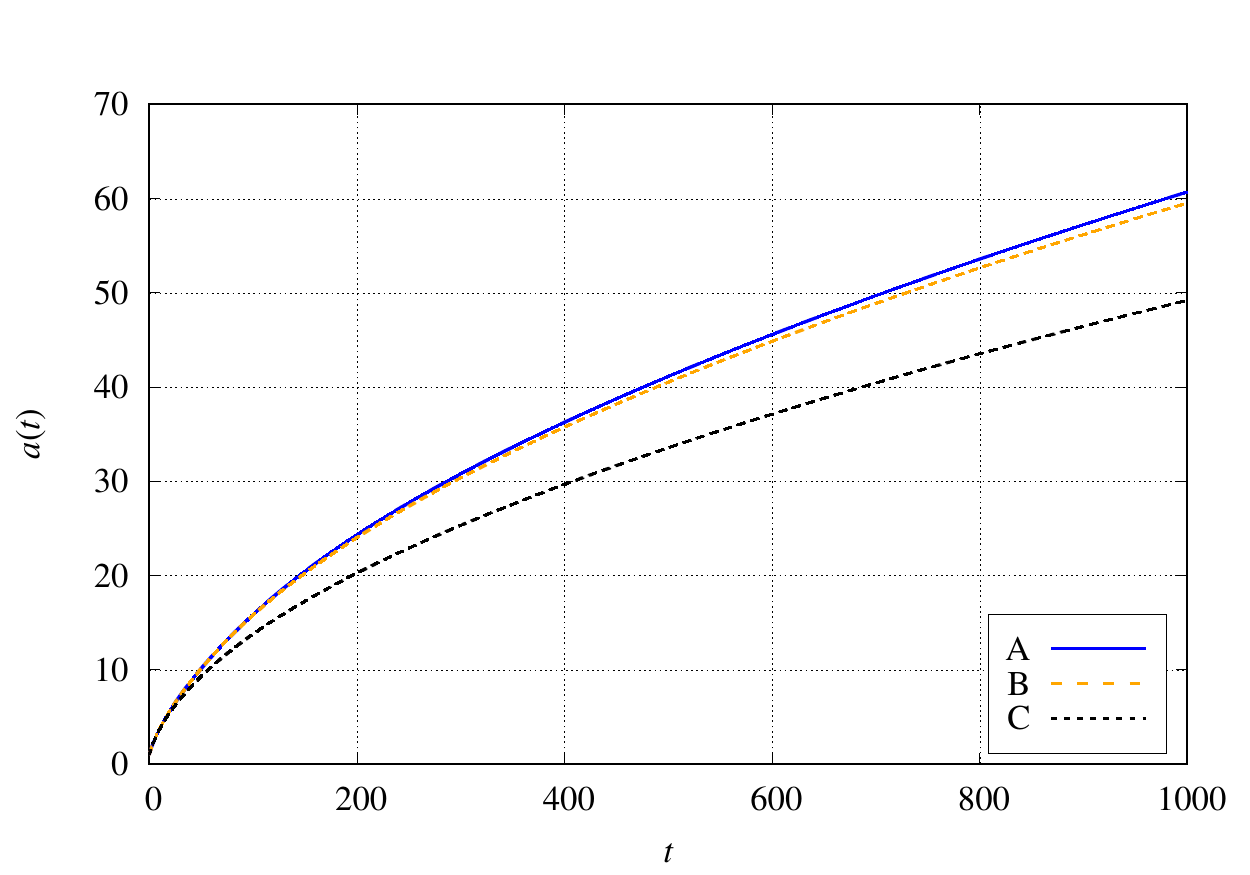}
\includegraphics[scale=.61]{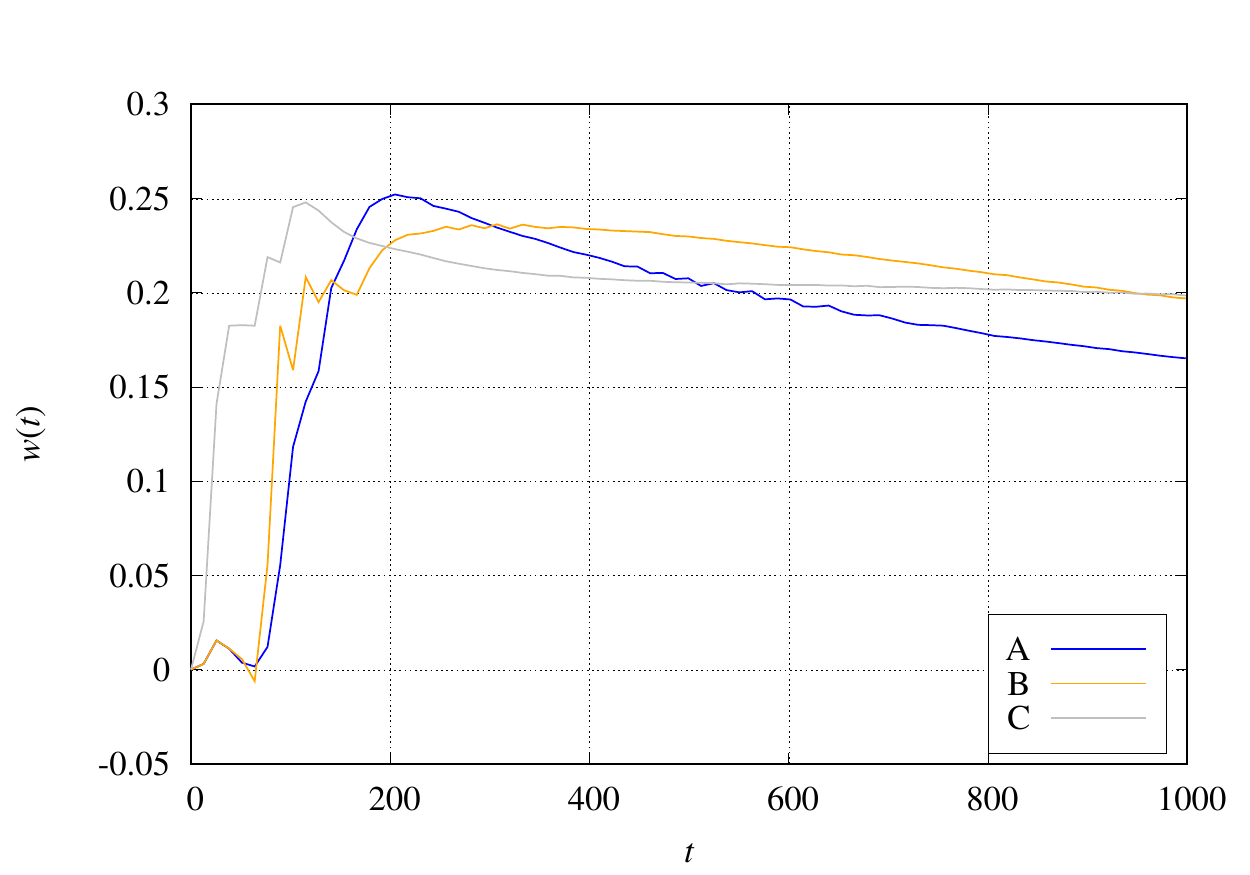}
\end{center}
\caption{\label{w_comp} Mean behavior of $a\left(t\right)$ and $w\left(t\right)$ for the models A, B and C.}
\end{figure}

We show in Fig. \ref{w_comp} the time evolutions of the mean behavior of effective equation of state and $w(t)$ (panel on the right) and scale factor $a(t)$ (panel on left) for all models that we have considered. The noticeable effect of the different interactions is to change the behavior of the scale factor growth and the effective equation of state on average. We conjecture this effect is related to the fact the 3-legs interaction acts anticipating the turbulent phase of the system and we are detailing in the next paragraph.
%
%

Analyzing the behavior of $w$, we can see all models start in $w \approx 0$. The model A has a peak in $w \approx 0.25$ and starts to tend to $w=0$ again. We conclude that models like this will never have a radiation domination era. On the other hand, the models B and C have a peak slightly smaller than $w \approx 0.25$ and they decay asymptotically to $w \approx 0.2$ that is an intermediate value between the matter and radiation domination eras. We conclude these models can evolve to a radiation era, but this peak has a different value to presented in \cite{dufaux2006}. We believe this difference might be associated with the artificial frequency cutoff due to numerical discretization. Dufaux et al. \cite{dufaux2006} have used a grid with $256^3$ points that give us a maximum wave number $k_{max} = 128\sqrt{3} \approx 221.7$. We have used a grid with de $64^3$ points that gives us a $k_{max} = 32\sqrt{3} \approx 55.4$. In the present numerical experiments, we notice that the cutoff frequency due to the discretization size can change the values of $w$. Apparently, there a value for $N$ where this effect becomes irrelevant. However, our results have the same qualitative behavior to those presented by Dufaux et al. \cite{dufaux2006}.

\subsection{Power spectra in spatial domain}

We use the same approach of the preceding paper \cite{crespo2019} to analyze the energy fluxes between different scales of the system. We present all physical and computational details in \cite{crespo2019}. The energy flows from larger scales to smaller scales through constant fluxes named energy cascades. The presence of these cascades is signalized by scale laws on the power spectra in the spatial domain, like the Kolmogorov's law 
\begin{equation}
P\left(k\right) \propto k^{-\gamma}, \label{kolmogorov_law}
\end{equation}

\noindent where $k$ is the wavenumber. 
We obtain the scale-law associated with the following quantities: the functions $\textrm{var}\left(\phi_p\right)$ and the rescaled energy density $\bar{\rho}$ 
\begin{equation}
\textrm{var}\left(\phi_p\right) = \left(\phi_p - \left\langle\phi_p\right\rangle\right)^2 \quad \mbox{and} \quad \bar{\rho} = \dfrac{a^4\rho}{\lambda\phi_0^4}.
\end{equation}

\begin{figure}[h!]
\includegraphics[scale=0.61]{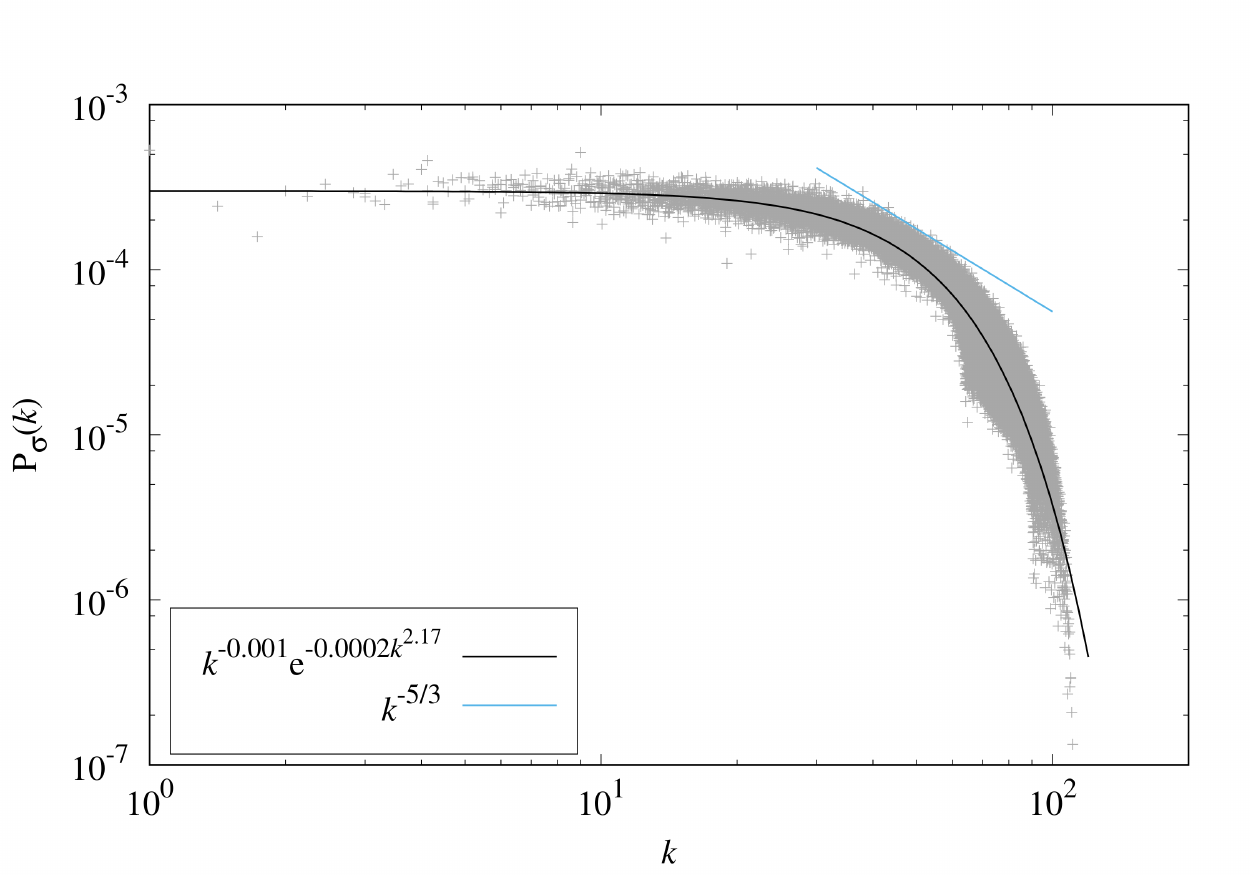}
\includegraphics[scale=0.61]{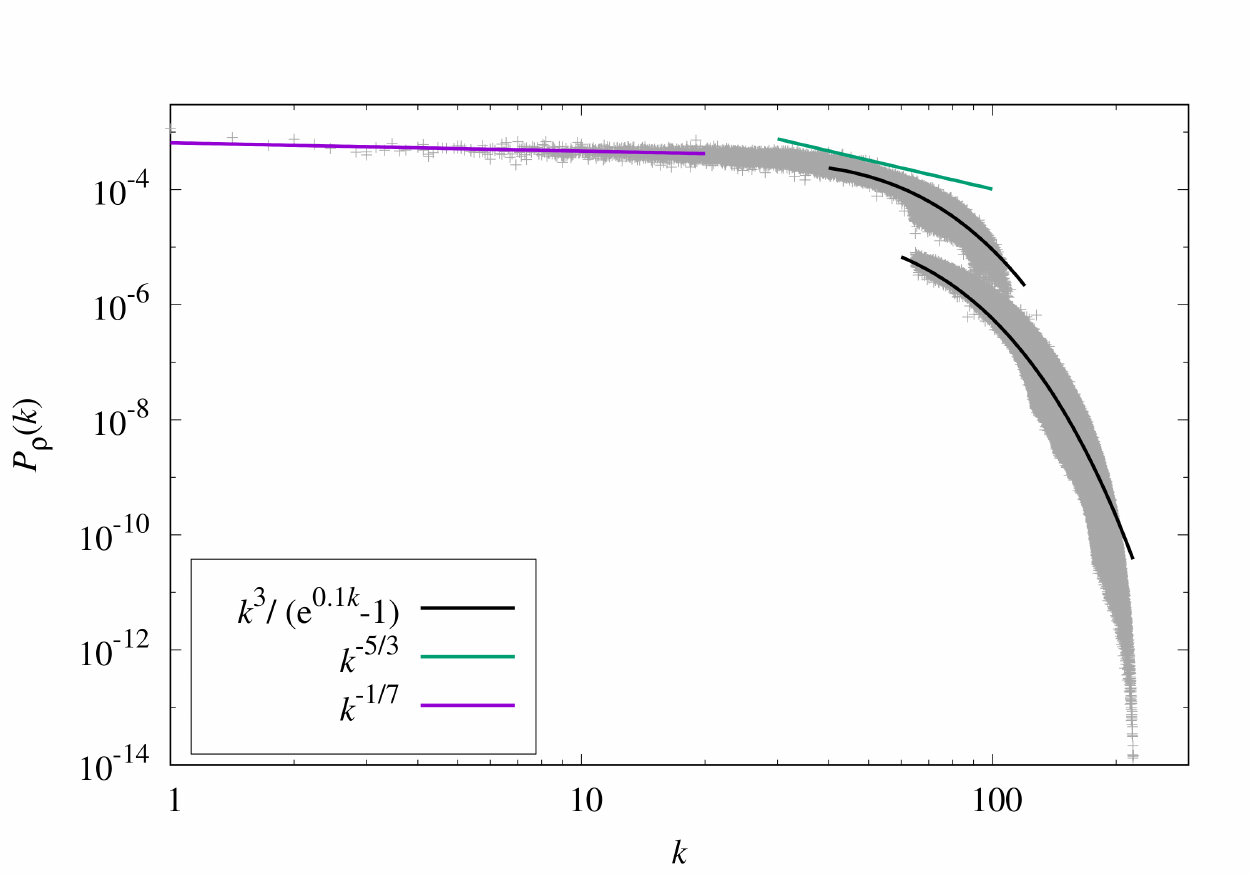}
\caption{\label{ps_0a} Power spectra in spatial domain for model A.}
\end{figure}
\begin{figure}[h!]
\includegraphics[scale=0.61]{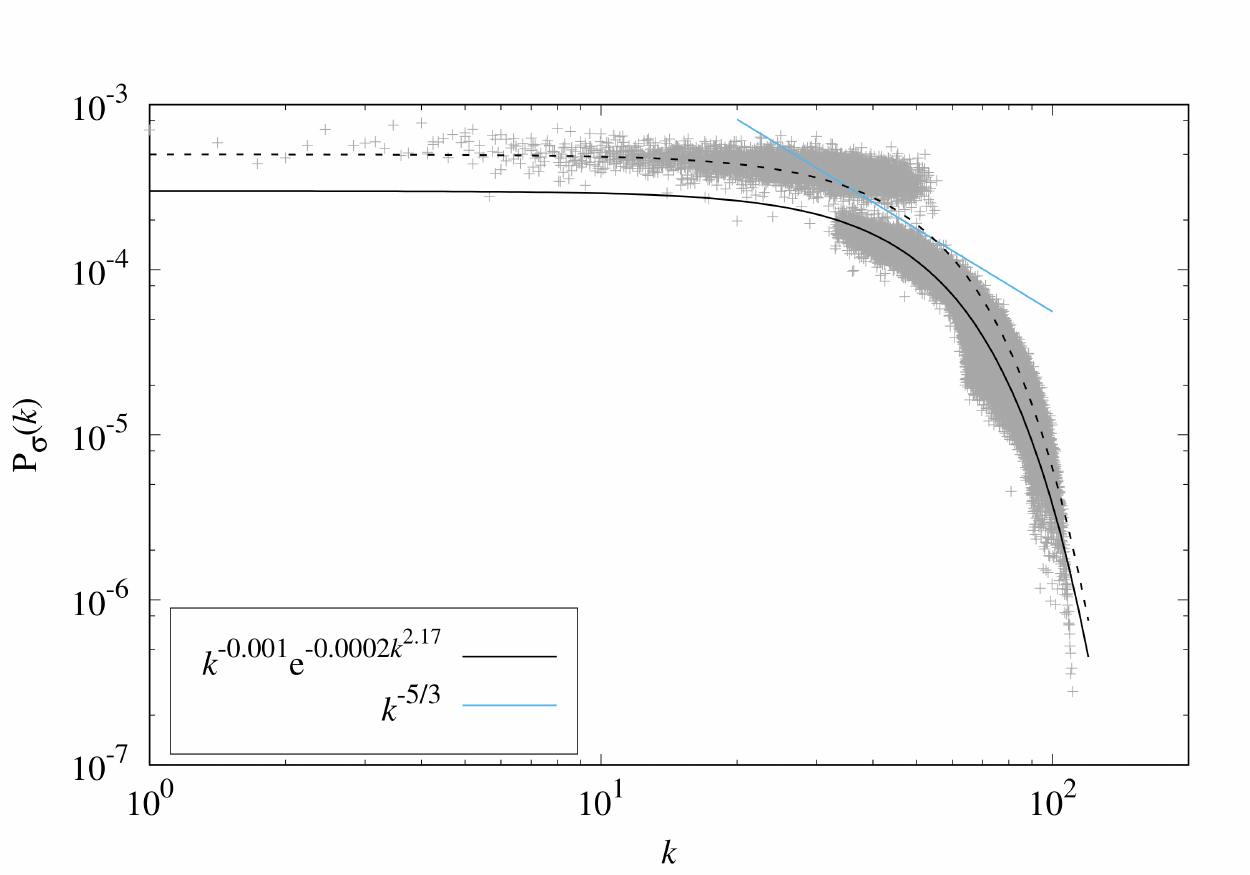}
\includegraphics[scale=0.61]{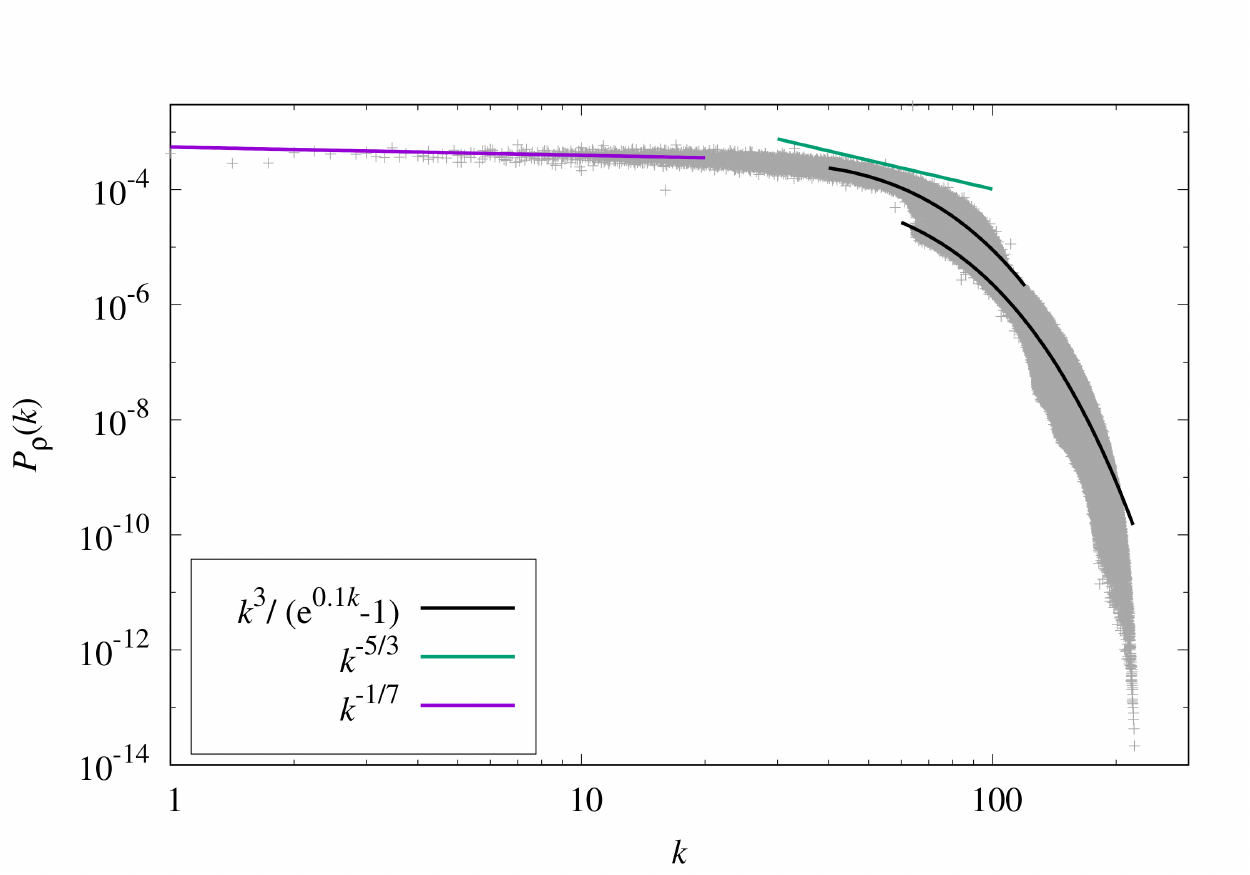}
\caption{\label{ps_01a} Power spectra in spatial domain for model B.}
\end{figure}
\begin{figure}[h!]
\includegraphics[scale=0.61]{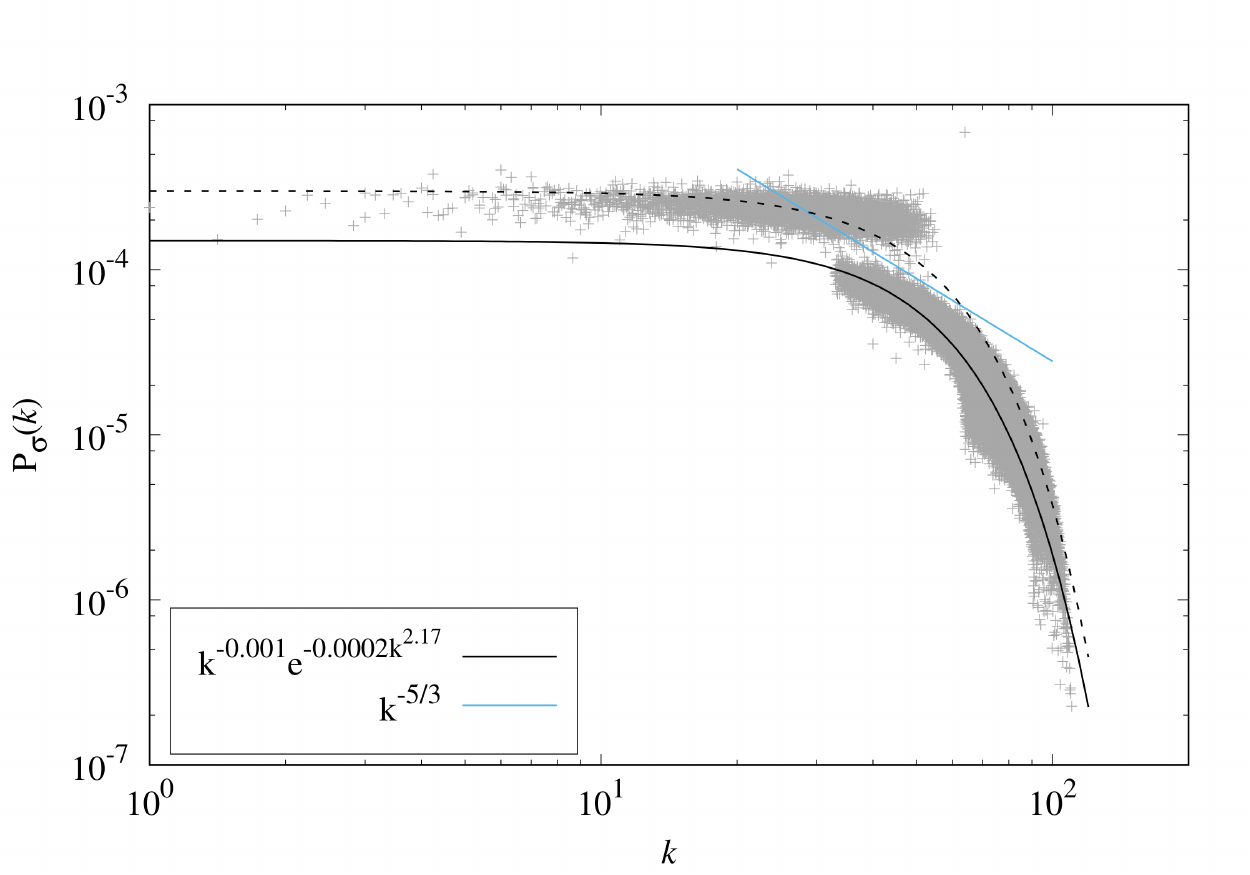}
\includegraphics[scale=0.61]{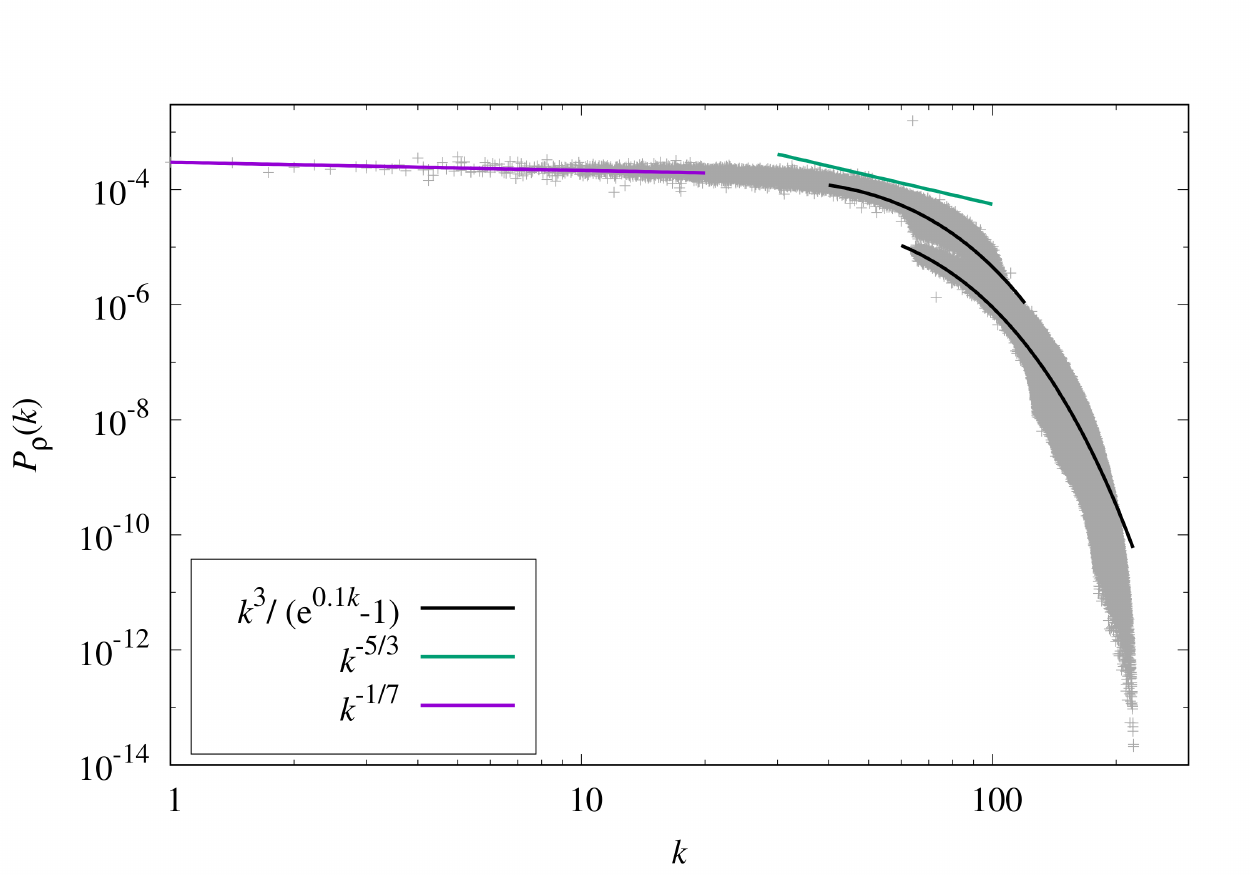}
\caption{\label{ps_1a} Power spectra in spatial domain for model C.}
\end{figure}

The aspects of wave turbulence imprinted in the power spectra evaluated at the late nonlinear stages of preheating seem to be universal since their structure for the two-fields models are similar to those of one field model \cite{crespo2019}. We show in Figs. \ref{ps_0a} - \ref{ps_1a} the power spectra of $\textrm{var}\left(\phi_p\right)$ (left) and $\bar{\rho}$ (right) evaluated at $\tau \approx 1000$ for models A, B and C, respectively. We noticed that all models have the scale law  
\begin{equation}
P_\sigma\left(k\right) \propto k^{-0.001}\expt^{-2.0 \times 10^{-4} k^{2.17}}
\end{equation}
\noindent for the power spectrum of $\textrm{var}\left(\phi_p\right)$. This scaling law occurs in magnetohydrodynamic turbulence \cite{terry09} and also in the case of a scalar field nonminimally coupled to the curvature \cite{crespo2019}. 

We have found a similar structure for the power spectrum of the energy density in compared with the case of a single field \cite{crespo2019}. For $k \lesssim 20$, all the spectra are described by the power-law  (\ref{kolmogorov_law}) with $\gamma \simeq 1/7$, whereas for $k > 40$ we were able to identify two distinct components described by the following one-parameter function
\begin{equation}
P_\rho\left(k\right) \propto \dfrac{k^3}{\expt^{bk}-1},\label{ps_rho}
\end{equation}

\noindent where $b \approx 0.1$ is independent of the model under consideration. We identify this scaling law as the Planck's law that describes the radiation spectrum of a black body. 


As in the case of the nonminimal scalar field model \cite{crespo2019}, the scaling law (\ref{ps_rho}) indicates the presence of the energy cascade that drives the system to a typical radiation energy distribution. We remark that the type of interaction (3 or 4-legs) does not alter the nature of the energy distribution despite the slight changes in the power spectrum as one can see in Figs. \ref{ps_0a} - \ref{ps_1a}. Another aspect denoting the cascade of energy is the presence of an intermediate range that can be described approximately by the power law (cf. eq. \ref{kolmogorov_law}) with $\gamma \approx 5/3$ (blue and green lines) that reveals the so-called inertial range. In the present simulations, we found the inertial range in the around the interval $30 \lesssim k \lesssim 80$.

Here again, we found strtucures shared with 3D numerical simulations of Bose-Einstein condensation (BEC) phenomenon within the nonlinear Schr{\"o}dinger (NLS) cited in \cite{nazarenko}. We also have noticed that $P_\phi\left(k\right)$ shows two components and the division between these components occurs in $k \approx 50$. In the particular case of our simulations, we have used the grid resolution $N=64$ and the maximum value for $k$ is $k_{max} = 32\sqrt{3} \approx 55.4$. We found a similar split in the 1-field nonminimally coupled models \cite{crespo2019}.

\subsection{Power spectra in time domain}

\begin{figure}[h]
\begin{center}
\includegraphics[scale=0.7]{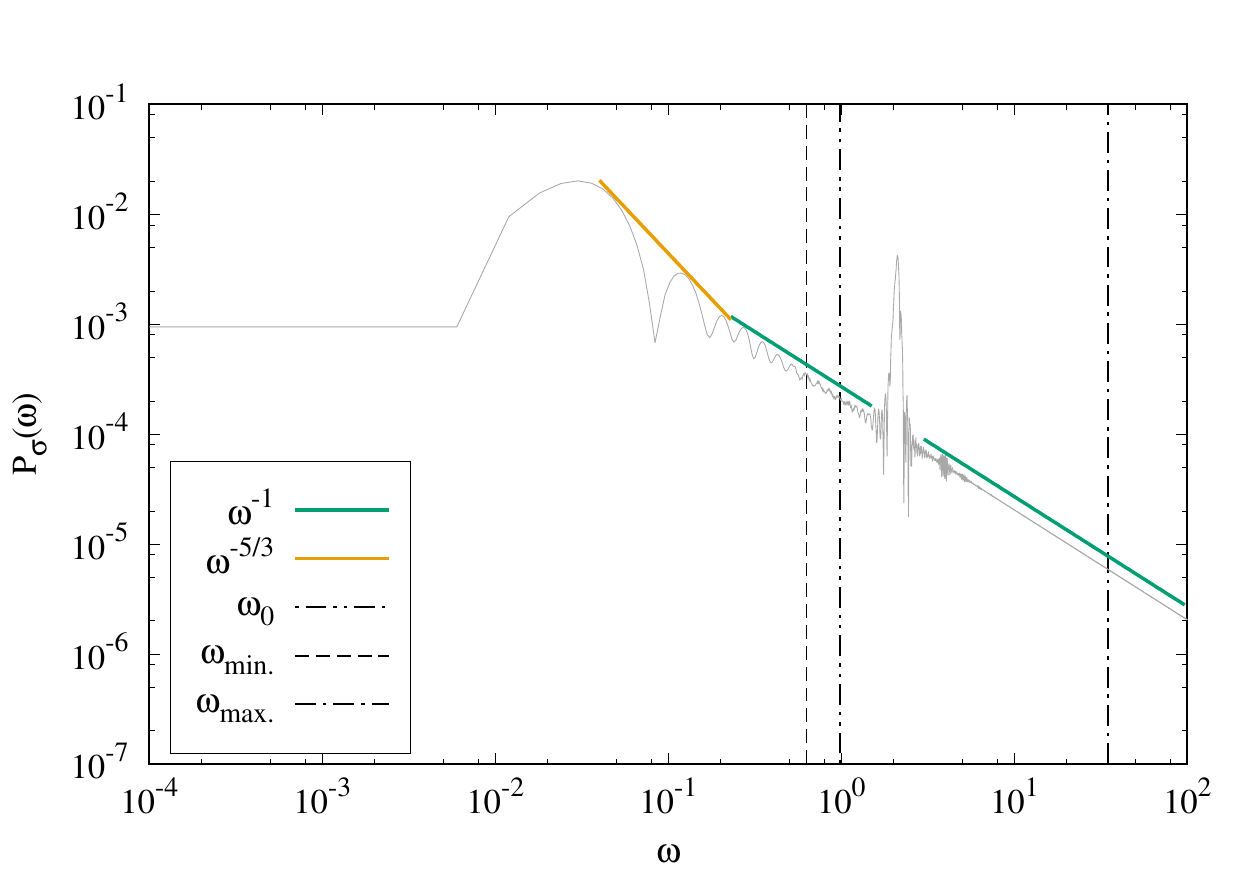}

\includegraphics[scale=0.7]{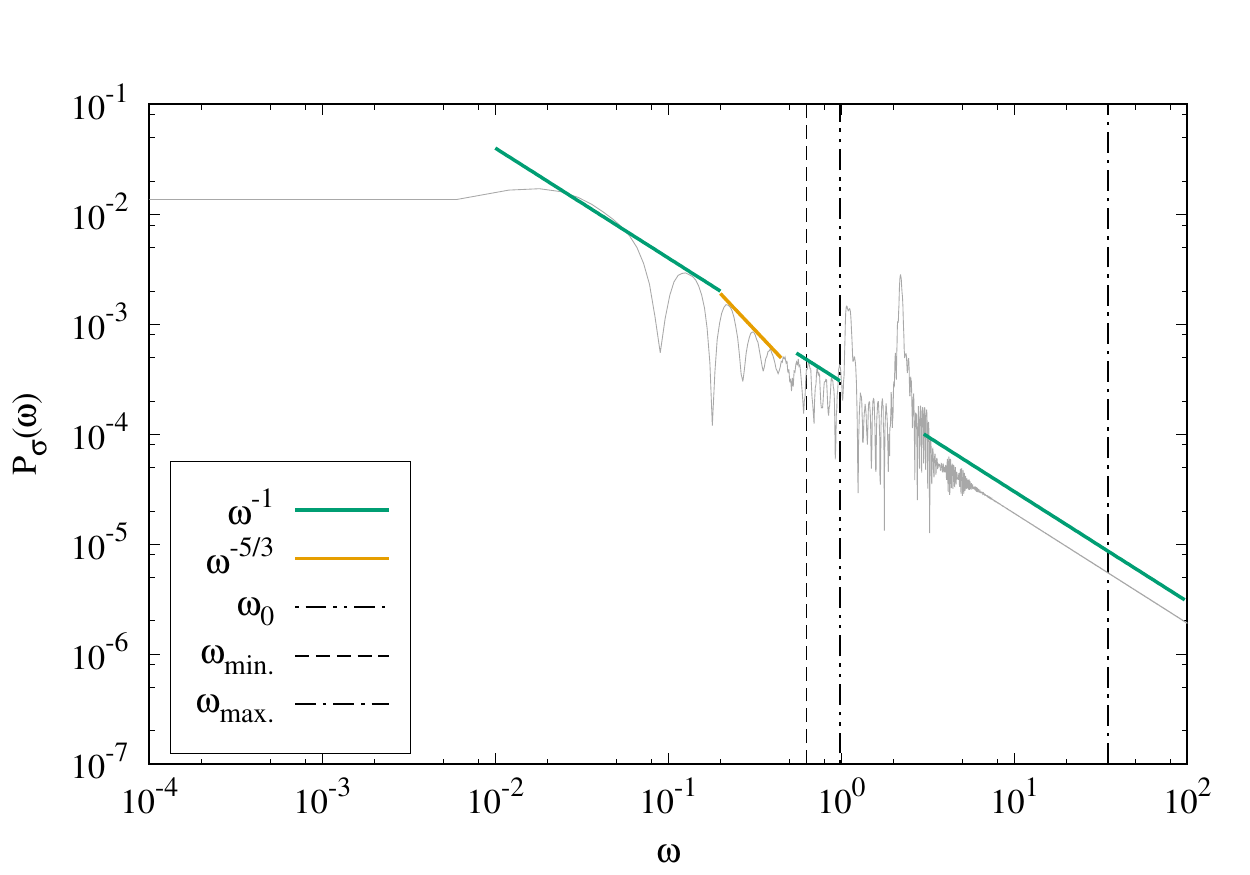}

\includegraphics[scale=0.7]{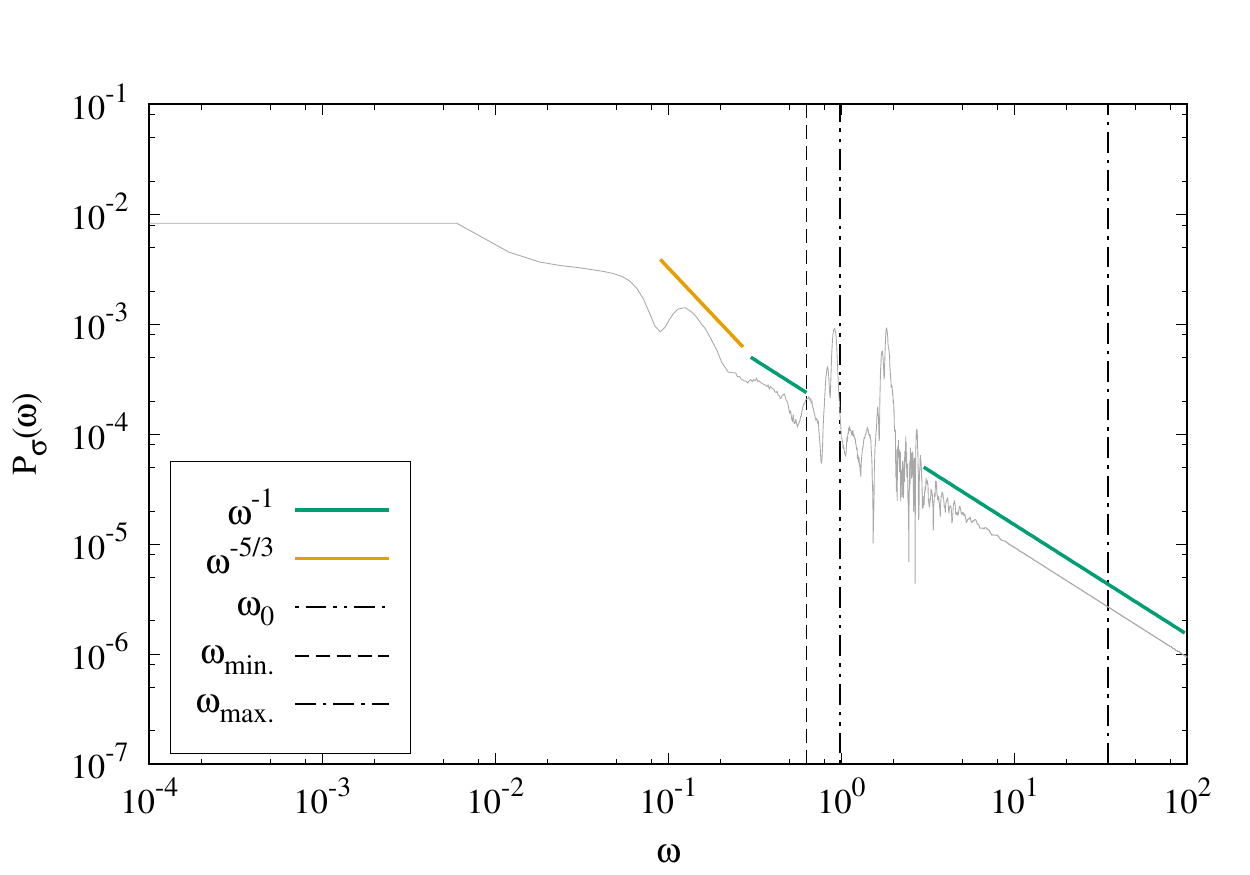}
\end{center}
\caption{\label{vps_0}Power spectra in time domain for models A, B and C.}
\end{figure}

We proceed exhibiting the power spectra of the variance, $\sigma_\phi^2$, in the time or frequency domain to obtain insights about the system energy flux and consequently about particle production in the late nonlinear stages of preheating. We use the same method presented in Ref. \cite{crespo2019}. Then, Fig. \ref{vps_0} shows the power spectra corresponding to the models A, B, and C, respectively. The feature shared by all power spectra is the presence of the Kolmogorov-Zakharov power law described by
\begin{equation}
P\left(\omega\right) \propto \omega^{-\gamma},
\end{equation}
\noindent with $\gamma$ assuming distinct values such as $\gamma \approx 1, 5/3$ depending on the ranges of $\omega$. As we have mentioned before, the presence of the above power-law is a signature of any turbulent system and the physical meaning is the same that we described in \cite{crespo2019}.

We noted that the exponent $\gamma = 1$ is independent of the 3 or 4 legs interaction present in the potential. In fact, this exponent fits almost all power spectra. This behavior occurs in generalized quantum models of Bose-Einstein gas \cite{nazarenko} and we found the same feature in the case of the one-field models of Ref. \cite{crespo2019}. In particular, for these models, the homogeneous mode frequency and oscillation period are, respectively, $\omega_{0} = 2 \pi/ \Pi_{\phi}$ and $\Pi_{\phi} = 6.4$; the boundary frequencies of spectral expansion are $\omega_{min.} = 2 \pi/ L_p$ and $\omega_{max.} = 64 \pi \sqrt{3} / L_{p}$.

Another aspect shared for all models, mainly A and B, is the presence of power-law related to stationary fluxes out of the frequency interval in spectral expansion. We believe this can be related with a shift effect in the power spectra of two interactng fields due to the effective mass of particles reported in \cite{micha04}.

\section{Energy distribution temperature and Universe thermalization}


We adopt the same procedure as in the Refs. \cite{henrique13,crespo2019} to estimate the temperature associated with the spectrum of the energy density in the wavenumber. As discussed, we identify an energy distribution typical of a black body satisfying a Planck-like law (33) (and Eq. (30) of the previous paper). We restore the physical variables in the exponential argument, since $bk = \hbar c k_{phys}/k_{B}T$, $k_{phys}$ is the physical \emph{momentum}, $k_{B}$ é is the Boltzmann's constant and $T$ is the temperature of the distribution. For all two interacting fields models, we have $k = L_p k_{phys} /m$ and $m \approx 10^{-6} m_{pl}$, we obtain
\begin{equation}
T = \dfrac{\hbar c m}{k_{B}bL_p} = 2.467 \times 10^{12} \mbox{ GeV}.
\end{equation}

\noindent We remark that this value is the same for all models under consideration and it is consistent with the upper bound of the reheating temperature of $10^{15}\,\mathrm{GeV}$ as pointed out in Ref. \cite{allahverdi10}. The above reheating temperature is higher than those values estimated previously in the case of a single nonminimally coupled scalar field. However, whereas in the latter case an effective equation of state representing radiation is obtained, here only the models B and C seem to present a form of matter at the late stages of preheating. 

\section{Discussion}

In this work, we extended the previous study of preheating considering models with two interacting scalar fields and taking into account the effect of backreaction of the field fluctuations in the expansion of the Universe as well as producing an effective equation of state. The main goal was to exhibit the aspects of wave turbulence in the late stages of preheating through the power spectra of the scalar field variance and the energy density. 

It is worth mentioning that in both study cases, namely the preheating with one nonminimally coupled scalar field with the spacetime curvature \cite{crespo2019} and the present preheating with two-fields, the results are similar as far as the power spectra are concerned. It reveals the universality of the turbulent regime once it is fully established. We found that in some intervals of the power spectra are described by power-laws $k^{-\gamma}$ and $\omega^{-\gamma}$, with $k$ and $\omega$ representing the wavenumber and the frequency, respectively, as a consequence of the energy transfer from the homogeneous inflaton field to the inhomogeneous fluctuations in the turbulent phase.


The next step for the present investigation is to examine the role of wave turbulence in the late stages of preheating taking into account the excitations of the gravitational fluctuations. It has been shown that it is possible that the perturbations of the metric can grow exponentially during the initial phase of preheating \cite{finelli99,basset99}. We may consider the case of two-field model as well as the single-field nonminimally coupled to curvature.

\section*{Acknowledgments}

J. A. Crespo acknowledges the financial support of the Brazilian agency Funda\c c\~ao Carlos Chagas Filho de Amparo \`a Pesquisa do Estado do Rio de Janeiro (FAPERJ). H. P. de Oliveira thanks to Conselho Nacional de Desenvolvimento Cient\'ifico e Tecnol\'ogico (CNPq) and Funda\c c\~ao Carlos Chagas Filho de Amparo \`a Pesquisa do Estado do Rio de Janeiro (FAPERJ) (Grant No. E-26/202.998/518 2016 Bolsas de Bancada de Projetos (BBP)).

\section*{References}

\bibliography{ref}

\providecommand{\newblock}{}
\begin{thebibliography}{10}
\expandafter\ifx\csname url\endcsname\relax
  \def\url#1{{\tt #1}}\fi
\expandafter\ifx\csname urlprefix\endcsname\relax\def\urlprefix{URL }\fi
\providecommand{\eprint}[2][]{\url{#2}}

\bibitem{henrique13}
Crespo J~A and de~Oliveira H~P 2014 {\em Journal of Cosmology and Astroparticle
  Physics\/} {\bf 2014} 006

\bibitem{crespo2019}
Crespo J~A and de~Oliveira H~P 2019  (\textit{Preprint} \eprint{1905.13647})

\bibitem{planck13}
{Planck Collaboration}, Ade P~A~R, Aghanim N, Armitage-Caplan C, Arnaud M,
  Ashdown M, Atrio-Barandela F, Aumont J, Baccigalupi C, Banday A~J and et~al
  2013 {\em ArXiv e-prints\/}

\bibitem{planck18}
Akrami Y {\em et~al.\/} (Planck) 2018  (\textit{Preprint} \eprint{1807.06211})

\bibitem{micha04}
Micha R and Tkachev I~I 2004 {\em Phys. Rev. D\/} {\bf 70}(4) 043538

\bibitem{dufaux2006}
Dufaux J~F, Felder G~N, Kofman L, Peloso M and Podolsky D 2006 {\em Journal of
  Cosmology and Astroparticle Physics\/} {\bf 2006} 006

\bibitem{orszag}
Gottlieb D and Orszag S 1977 {\em Numerical Analysis of Spectral Methods\/}
  (Society for Industrial and Applied Mathematics)

\bibitem{kofman94}
Kofman L, Linde A and Starobinsky A~A 1994 {\em Phys. Rev. Lett.\/} {\bf
  73}(24) 3195--3198

\bibitem{prokopec97}
Prokopec T and Roos T~G 1997 {\em Phys. Rev. D\/} {\bf 55}(6) 3768--3775

\bibitem{greene97}
Greene P~B, Kofman L, Linde A and Starobinsky A~A 1997 {\em Phys. Rev. D\/}
  {\bf 56}(10) 6175--6192

\bibitem{kofman97}
Kofman L, Linde A and Starobinsky A~A 1997 {\em Phys. Rev. D\/} {\bf 56}(6)
  3258--3295

\bibitem{kofman98}
Kofman L 1998 {\em ArXiv Astrophysics e-prints\/}

\bibitem{podolski06}
Podolsky D, Felder G~N, Kofman L and Peloso M 2006 {\em Phys. Rev. D\/} {\bf
  73}(2) 023501

\bibitem{kofman2001}
Kofman L 2001 {Tachyonic Preheating} {\em Particles, Strings, and Cosmology\/}
  ed Frampton P and Ng J p 167 (\textit{Preprint} \eprint{hep-ph/0107280})

\bibitem{planck2016}
{Planck Collaboration}, Ad P~A~R, Aghanim N, Arnaud M, Arroja F, Ashdown M,
  Aumont J, Baccigalupi C, Ballardini M, Banday A~J and et~al 2016 {\em
  Astronomy and Astrophysics\/} {\bf 594} A20 (\textit{Preprint}
  \eprint{1502.02114})

\bibitem{harigaya2016}
Harigaya K, Ibe M, Kawasaki M and Yanagida T~T 2016 {\em Physics Letters B\/}
  {\bf 756} 113 -- 117

\bibitem{terry09}
Terry P~W and Tangri V 2009 {\em Physics of Plasmas\/} {\bf 16} 082305

\bibitem{nazarenko}
Nazarenko S 2011 {\em Wave turbulence\/} (Berlin: Springer)

\bibitem{allahverdi10}
Allahverdi R, Brandenberger R, Cyr-Racine F~Y and Mazumdar A 2010 {\em Annual
  Review of Nuclear and Particle Science\/} {\bf 60} 27--51 (\textit{Preprint}
  \eprint{1001.2600})

\bibitem{finelli99}
Finelli F and Brandenberger R 1999 {\em Phys. Rev. Lett.\/} {\bf 82}(7)
  1362--1365

\bibitem{basset99}
Bassett B~A, Tamburini F, Kaiser D~I and Maartens R 1999 {\em Nuclear Physics
  B\/} {\bf 561} 188 -- 240

\end{thebibliography}

\end{document}